\documentclass[a4paper,fleqn,usenatbib]{mnras}

\usepackage[T1]{fontenc}
\usepackage{ae,aecompl}

\usepackage{graphicx}
\usepackage{upgreek}
\usepackage{amssymb}

\newcommand{\acronym}[1]{\textsc{#1}}
\newcommand{\software}[1]{\textsc{#1}}
\newcommand{\code}[1]{\texttt{#1}}
\newcommand{\sofia}[0]{\software{SoFiA}}
\newcommand{\wallaby}[0]{WALLABY}
\newcommand{\unit}[1]{\ensuremath{\mathrm{#1}}}
\newcommand{\hi}[0]{\ensuremath{\textrm{H\,I}}}

\title[SoFiA~2 source finding pipeline]{SoFiA~2 -- An automated, parallel \hi{} source finding pipeline for the WALLABY survey}

\author[T.~Westmeier et al.]
	{T.~Westmeier,$^{1,2}$\thanks{E-mail: tobias.westmeier@uwa.edu.au (TW)}
	S.~Kitaeff,$^{1,3}$
	D.~Pallot,$^{1,3}$
	P.~Serra,$^{4}$
	J.~M.~van der Hulst,$^{5}$
	\newauthor
	R.~J.~Jurek,$^{6}$
	A.~Elagali,$^{7,2}$
	B.-Q.~For,$^{1,2}$
	D.~Kleiner,$^{4}$
	B.~S.~Koribalski,$^{7,8}$
	\newauthor
	K.~Lee-Waddell,$^{7}$
	J.~R.~Mould,$^{9}$
	T.~N.~Reynolds$^{1,2}$
	J.~Rhee,$^{1,2}$
	\newauthor
	and L.~Staveley-Smith$^{1,2}$
	\\
	$^{1}$International Centre for Radio Astronomy Research (ICRAR), M468, The University of Western Australia, 35~Stirling Highway,\\ Crawley WA 6009, Australia\\
	$^{2}$ARC Centre of Excellence for All Sky Astrophysics in 3~Dimensions (ASTRO~3D)\\
	$^{3}$Australian SKA Regional Centre (AusSRC)\\
	$^{4}$INAF -- Osservatorio Astronomico di Cagliari, Via della Scienza~5, 09047 Selargius, CA, Italy\\
	$^{5}$Kapteyn Astronomical Institute, University of Groningen, Postbus~800, 9700AV Groningen, The Netherlands\\
	$^{6}$Department of Physics, School of Mathematics and Physics, The University of Queensland, Brisbane QLD 4072, Australia\\
	$^{7}$CSIRO Astronomy and Space Science, Australia Telescope National Facility, PO Box 76, Epping NSW 1710, Australia\\
	$^{8}$Western Sydney University, Locked Bag 1797, Penrith NSW 2751, Australia\\
	$^{9}$CAS, Swinburne University, PO Box 218, Hawthorn VIC 3122, Australia
}

\date{Accepted XXX. Received YYY; in original form ZZZ}
\pubyear{2020}

\begin{document}
	\label{firstpage}
	\pagerange{\pageref{firstpage}--\pageref{lastpage}}
	\maketitle
	
	\begin{abstract}
		We present \sofia{}~2, the fully automated 3D source finding pipeline for the \wallaby{} extragalactic \hi{} survey with the Australian SKA Pathfinder (ASKAP). \sofia{}~2 is a reimplementation of parts of the original \sofia{} pipeline in the C programming language and makes use of OpenMP for multi-threading of the most time-critical algorithms. In addition, we have developed a parallel framework called \mbox{\sofia{}-X} that allows the processing of large data cubes to be split across multiple computing nodes. As a result of these efforts, \sofia{}~2 is substantially faster and comes with a much reduced memory footprint compared to its predecessor, thus allowing the large \wallaby{} data volumes of hundreds of gigabytes of imaging data per epoch to be processed in real-time. The source code has been made publicly available to the entire community under an open-source licence. Performance tests using mock galaxies injected into genuine ASKAP data suggest that in the absence of significant imaging artefacts \sofia{}~2 is capable of achieving near-100\% completeness and reliability above an integrated signal-to-noise ratio of about 5--6. We also demonstrate that \sofia{}~2 generally recovers the location, integrated flux and $w_{20}$ line width of galaxies with high accuracy. Other parameters, including the peak flux density and $w_{50}$ line width, are more strongly biased due to the influence of the noise on the measurement. In addition, very faint galaxies below an integrated signal-to-noise ratio of about 10 may get broken up into multiple components, thus requiring a strategy to identify fragmented sources and ensure that they do not affect the integrity of any scientific analysis based on the \sofia{}~2 output.
	\end{abstract}
	
	
	\section{Introduction}
	
	Several precursors and pathfinders to the Square Kilometre Array (SKA; \citealt{Dewdney2009}) have recently begun taking early data, including the Australian SKA Pathfinder (ASKAP; \citealt{Hotan2021}), MeerKAT (\citealt{Jonas2009}; \citealt{Camilo2018}) and the APERture Tile In Focus (APERTIF; \citealt{Verheijen2008}). One of the major surveys to be carried out with ASKAP is the Widefield ASKAP L-band Legacy All-sky Blind Survey (\wallaby{}; \citealt{Koribalski2020}) which is expected to image the entire sky south of a declination of $+30\degr$ in the 21-cm line emission of neutral hydrogen (\hi{}). \wallaby{} is expected to produce about 1~petabyte of \hi{} imaging data and detect the \hi{} emission from approximately half a million galaxies out to a redshift of $z \approx 0.26$.
	
	Given the unprecedented amount of imaging data anticipated from extragalactic \hi{} surveys such as \wallaby{}, the detection and characterisation of galaxies will need to occur in a fully automated fashion with minimal manual intervention using dedicated spectral-line source finding software such as \software{Duchamp} \citep{Whiting2012}, \software{S\'{e}lavy} \citep{Whiting2012b} or the Source Finding Application (\sofia{}; \citealt{Serra2015}). The performance of several source finding packages and algorithms, and their suitability for \hi{} source finding, was tested by \citet{Popping2012}.
	
	\sofia{} has been specifically developed for the purpose of detecting galaxies in extragalactic \hi{} surveys like \wallaby{}. It features several powerful algorithms that have greatly improved the quality and accuracy of \hi{} source finding (e.g., \citealt{deBlok2018,For2019,BlueBird2020}), including the \emph{smooth and clip} (S+C) algorithm and a new method for automatically identifying and removing unreliable detections \citep{Serra2012a}.
	
	While \sofia{}'s algorithms are well-suited to detecting galaxies in \wallaby{} data, there are several shortcomings to the current implementation of \sofia{} that make its application to large data volumes challenging, most notably its large memory footprint and comparatively low speed. In order to address these and other issues, we have reimplemented the most powerful algorithms of \sofia{} in the C programming language in order to create a much faster and more memory-efficient pipeline called \sofia{}~2. In addition to improvements in the implementation, we employ techniques such as multi-threading and parallelisation to significantly speed up the processing of large data volumes, allowing \sofia{}~2 to process a single 800~GB \wallaby{} data cube in a matter of minutes rather than hours on a modest number of computing nodes.
	
	Here we introduce \sofia{}~2 and present its basic performance measures based on tests with both mock data and real \wallaby{} \hi{} data. A brief overview of the implementation of \sofia{}~2 and the main improvements compared to the original \sofia{} (hereafter referred to as \sofia{}~1 for clarity) is presented in Section~\ref{sect_implementation}, followed in Section~\ref{sect_parallel} by a description of the parallel framework named \mbox{\sofia{}-X}. Section~\ref{sect_algorithms} outlines some of the core algorithms employed by \sofia{}~2. In Section~\ref{sect_speed_and_memory} we present tests of the speed and memory usage of the pipeline, while the performance of \sofia{}~2 with respect to completeness, reliability and parameterisation accuracy is presented in Section~\ref{sect_performance} based on tests with mock galaxies injected into genuine ASKAP \hi{} data. Initial results from the processing of WALLABY pilot survey data with \sofia{}~2 are presented and discussed in Section~\ref{sect_pilot} followed by a summary and more general discussion in Section~\ref{sect_summary}.

	\section{Implementation}
	\label{sect_implementation}
	
	\sofia{}~2 is a reimplementation of the core algorithms of the original \sofia{}~1 pipeline, most notably automatic flagging, noise normalisation, S+C source finder and reliability estimation \citep{Serra2012a}. Some of these algorithms have also been simplified and improved compared to the original \sofia{} implementation, while other, less widely-used algorithms and features have not yet been ported, including the 2D--1D wavelet filter \citep{Floeer2012}, the CNHI source finder \citep{Jurek2012}, position--velocity diagrams, automatic scaling of the reliability measurement kernel and more sophisticated options for controlling the shape and size of the smoothing kernels in the S+C finder.
	
	Unlike the original \sofia{}~1 pipeline, which was written in a combination of Python, Cython and C++, \sofia{}~2 is written entirely in the C programming language, more specifically using the C99 standard (formally known as ISO/IEC~9899:1999). This has resulted in several substantial improvements:
	\begin{enumerate}
		\item The conversion to C alone has resulted in a modest improvement in speed and a substantial improvement in memory usage as compared to \sofia{}~1, even without multi-threading or parallelisation (see Section~\ref{sect_speed_and_memory} for details).
		\item As time-critical algorithms can be coded directly in C, the number of external library dependencies has been reduced to just one (\acronym{wcslib}; \citealt{Calabretta2011}), making \sofia{}~2 much easier to install and maintain.
		\item Thanks to the native support of OpenMP by commonly used C compilers, most time-critical algorithms in \sofia{}~2 have been fully multi-threaded, resulting in additional significant gains in speed on multi-core architectures.
	\end{enumerate}
	The choice of OpenMP \citep{Dagum1998} for multi-threading was made to keep the code as simple and easy to use as possible. OpenMP is natively supported by many compilers and therefore does not impose any additional dependencies. In addition, it can be adopted with just minimal modifications to the source code and can easily be disabled if necessary.
	
	Like its predecessor, the source code of \sofia{}~2 has been publicly released on GitHub\footnote{\url{https://github.com/SoFiA-Admin/SoFiA-2/}} under an open-source licence to make it freely available to the entire astronomy community. While the code has been optimised for compilation with the GNU Compiler Collection's \software{gcc} compiler,\footnote{\url{https://gcc.gnu.org/}} it should in principle compile with any C compiler that supports the C99 standard and is therefore expected to be able to run on a wide range of POSIX-compliant systems.
	
	As the main purpose of \sofia{}~2 is the automated detection and parameterisation of galaxies in the large volumes of data produced by \hi{} surveys such as \wallaby{}, we do not provide a graphical user interface as with \sofia{}~1. However, like its predecessor, \sofia{}~2 can be easily controlled through basic parameter files and comes with a robust set of default parameter settings that can serve as a starting point for establishing optimal parameter settings.

	\begin{figure*}
		\includegraphics[width=0.85\linewidth]{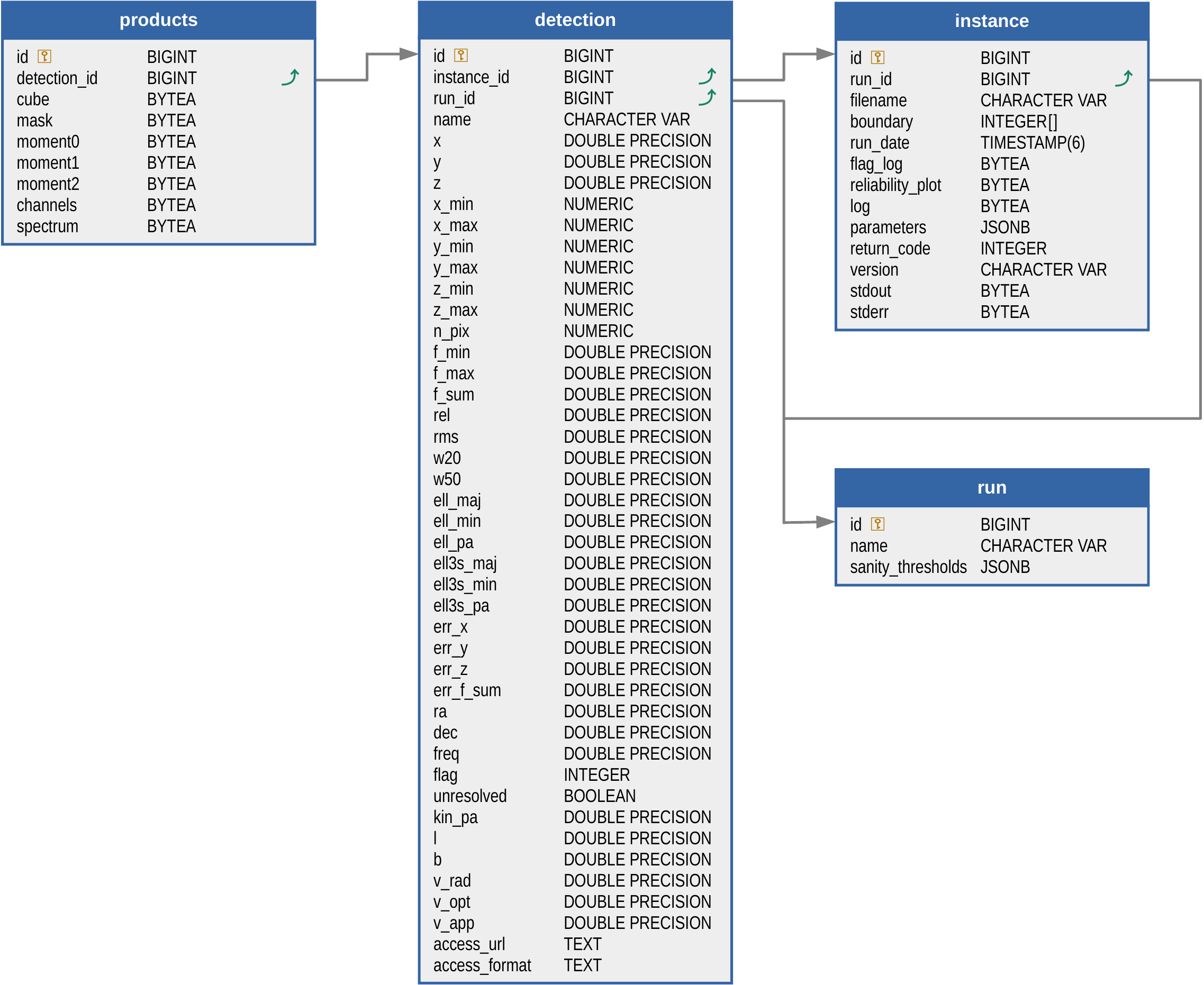}
		\caption{\mbox{\sofia{}-X} database schema showing the field names and data types in each of the four tables used. Each \emph{run} of \mbox{\sofia{}-X} spawns multiple parallel \emph{instances} of \sofia{}~2 that will typically operate on different regions of the data cube. The source parameters for each \emph{detection} will be written to a common table, with another table storing the image \emph{products} associated with each detection.}
		\label{fig_database_schema}
	\end{figure*}

	\section{Parallel framework}
	\label{sect_parallel}
	
	A single \wallaby{} spectral image cube is expected to be up to 800~GB in size, which immediately raises two issues. Firstly, it would take a significant amount of time to process such a large data volume on a single computing node even with multi-threading enabled. Secondly, up to 2~TB of memory would be required to load and process such a cube all at once (see Section~\ref{sect_speed_and_memory}). The only viable solution to these two problems would be to split up the data cube into multiple sub-regions each of which must be small enough to fit into the amount of memory available on a single node. The individual sub-regions can then be processed in parallel on separate computing nodes and the resulting catalogues be merged at the end of this process into a single source catalogue of the full region.
	
	\sofia{}~2 already has the capability to read arbitrary sections of a data cube. The merging of multiple outputs, however, would still need to be carried out manually, including the difficult task of resolving cases of duplicate detections in overlapping regions. In order to automate this entire process, including the merging of output from multiple, partially overlapping regions of a data cube, or from multiple runs of the source finder on the same region, we  have developed the \mbox{\sofia{}-X} framework.
	
	\mbox{\sofia{}-X} is a wrapper around \sofia{}~2 that can spawn multiple instances of the pipeline across different nodes of a computing cluster using the Slurm Workload Manager \citep{Yoo2003} and collate the output from all instances into a single, coherent source catalogue. The choice of Slurm was again driven by the idea of simplicity and ease of use, as Slurm is open-source and readily available on most supercomputers. The instances of \sofia{}~2 processing the regions require no interprocess communications and therefore can be executed as an array job in Slurm. The scripts can be easily changed if a different batch job scheduler is desired or available.
	
	Each instance of the source finding pipeline is provided with a configuration file that specifies a region of the data cube to be read and produces a catalogue of detections for that region in VOTable\footnote{\url{http://www.ivoa.net/documents/VOTable/}} format. There are no limitations on how these regions are defined, and it is entirely up to the user to choose suitably sized spatial and/or spectral regions with adequate overlap by using the \code{input.region} keyword in the \sofia~2 parameter file.
	
	Once the VOTable is created, it is then passed to a Python script that imports the catalogues and image products for each detected source into a database on a server. The location of the database server can be the same for all runs of \sofia{}~2 on a cluster, or the scripts can be pointed to the different locations of multiple databases. Each run and each instance of the programme are uniquely identified in the database. The \sofia{}~2 parameter settings of the runs are also recorded in the database as metadata to enable provenance of the data and retrospective reviews. The database schema is presented in Fig.~\ref{fig_database_schema}.
	
	During the import of each catalogue, a Python script attempts to automatically identify and remove duplicates in overlapping regions by checking if the positions and a few other observational parameters agree within certain thresholds specified by the user. If the heuristics of the automated decision making does not allow unambiguous resolution of duplicates, then the detections will be flagged for manual resolution in the database. For example, this could be the case if the positions of two sources agreed within the specified thresholds, but their fluxes did not.
	
	\mbox{\sofia{}-X} provides the user with a convenient web portal for inspecting the merged catalogue and manually resolving any flagged detections. The portal has a \emph{Table Access Protocol} (TAP)\footnote{\url{http://www.ivoa.net/documents/TAP/}} interface that allows users to directly connect through VO-compliant software, such as \software{topcat} \citep{Taylor2005}, or run ADQL\footnote{\url{http://www.ivoa.net/documents/ADQL/}} queries on the source database. The image products of individual sources are accessible through VO-compliant data links and can be directly visualised with software such as Aladin \citep{Bonnarel2011}.
	
	The source code of \mbox{\sofia{}-X} has been made publicly available and can be accessed on GitHub.\footnote{\url{https://github.com/AusSRC/SoFiAX/}} Because \mbox{\sofia{}-X} acts as a wrapper around \sofia{}~2, it can be installed independently and is not required for running just a single instance of \sofia{}~2.

	\section{Algorithms}
	\label{sect_algorithms}
	
	\sofia{}~2 largely uses the same algorithms as its predecessor. For that reason we will refrain from a detailed description of the individual algorithms and instead refer the reader to \citet{Serra2015} and references therein. Detailed information about the different algorithms and settings is also available from the official \sofia{}~2 user manual which can be obtained from the \sofia{}~2 GitHub wiki. Here we will give a brief overview of some of the core modules of \sofia{}~2 with a particular focus on algorithms that are relevant to the analysis presented in this paper or have since been added or changed. At the time of writing, a few useful features from \sofia{}~1 have not yet been incorporated into \sofia{}~2 (see Section~\ref{sect_implementation}) and are expected to be implemented in the near future.

	\subsection{Input and preconditioning}
	
	Like its predecessor, \sofia{}~2 currently only supports input data in the \emph{Flexible Image Transport System} (FITS) format \citep{Pence2010}. Data files must have 2--4 dimensions as long as the fourth axis has a size of no greater than one (e.g.\ Stokes~$I$). It is possible to read and process only a sub-region of a data cube. In addition to the actual data cube to be searched, the user can also specify an optional noise cube, weights cube or gain cube. These will be applied to the data cube prior to either source finding or source parameterisation, as appropriate.
	
	Users can manually specify a set of rectangular regions to be flagged prior to source finding. In addition, a new auto-flagging algorithm is available which will dynamically flag spatial pixels or spectral channels for which the noise deviates from the median noise across all pixels or channels by a user-specified multiple of the RMS as estimated from the more robust median absolute deviation. The auto-flagger is intended for the automatic flagging of corrupted data, including channels affected by radio-frequency interference or pixels containing residual continuum emission.
	
	If for some reason the noise level across the input data cube is not constant, then \sofia{}~2 will first need to normalise the noise level before being able to apply a constant threshold to the data for the purpose of source finding. This can be achieved by providing either a noise cube by which the data cube will be divided, or a weights cube (holding the inverse of the variance) by the square root of which the data cube will be multiplied. If neither of those are available, or there are residual noise variations not accounted for by the noise or weights cube, then \sofia{}~2 can offer to measure and divide by the local noise level in a running window.

	\subsection{Source finding}
	
	\sofia{}~2 offers two different source finding algorithms: a simple threshold finder and the \emph{smooth and clip} (S+C) finder. The threshold finder will simply apply an absolute or relative (to the noise) flux threshold to the data and is rarely useful unless the data have already been preconditioned before being read into \sofia{}~2. The most powerful and default source finding algorithm of \sofia{}~2 is the S+C finder which is described in detail in \citet{Serra2012b}. It essentially works by iteratively smoothing the data cube on multiple spatial and spectral scales to extract statistically significant emission above a user-specified detection threshold on each scale. The output from both source finding algorithms will be a binary mask of detected pixels.

	An important feature of \sofia{}~2 is its ability to pick up both positive and negative flux density values in excess of the source finding threshold. This strategy avoids creating a positive flux bias that would inevitably arise if only positive flux density values were added to the source mask. In addition, the resulting false detections with negative total flux can be used to estimate the statistical reliability of detections with positive flux (see Section~\ref{sect_reliability}).

	\subsection{Linking}
	
	The purpose of the linker is to combine the detected pixels in the binary mask from the source finder into individual, coherent detections. For this purpose the linker uses a simple friends-of-friends algorithm that links all pixels within a user-specified merging radius and assigns a unique identifier to each detection. This linking of pixels occurs both in the spatial plane and along the spectral axis, i.e.\ sources are effectively treated as three-dimensional collections of pixels. In addition, user-specified minimum and maximum size filters can be applied in the spatial and spectral dimension to remove very large or very small detections that are likely to be false detections due to noise peaks or large-scale artefacts in the data.

	\begin{figure*}
		\includegraphics[height=5.9cm]{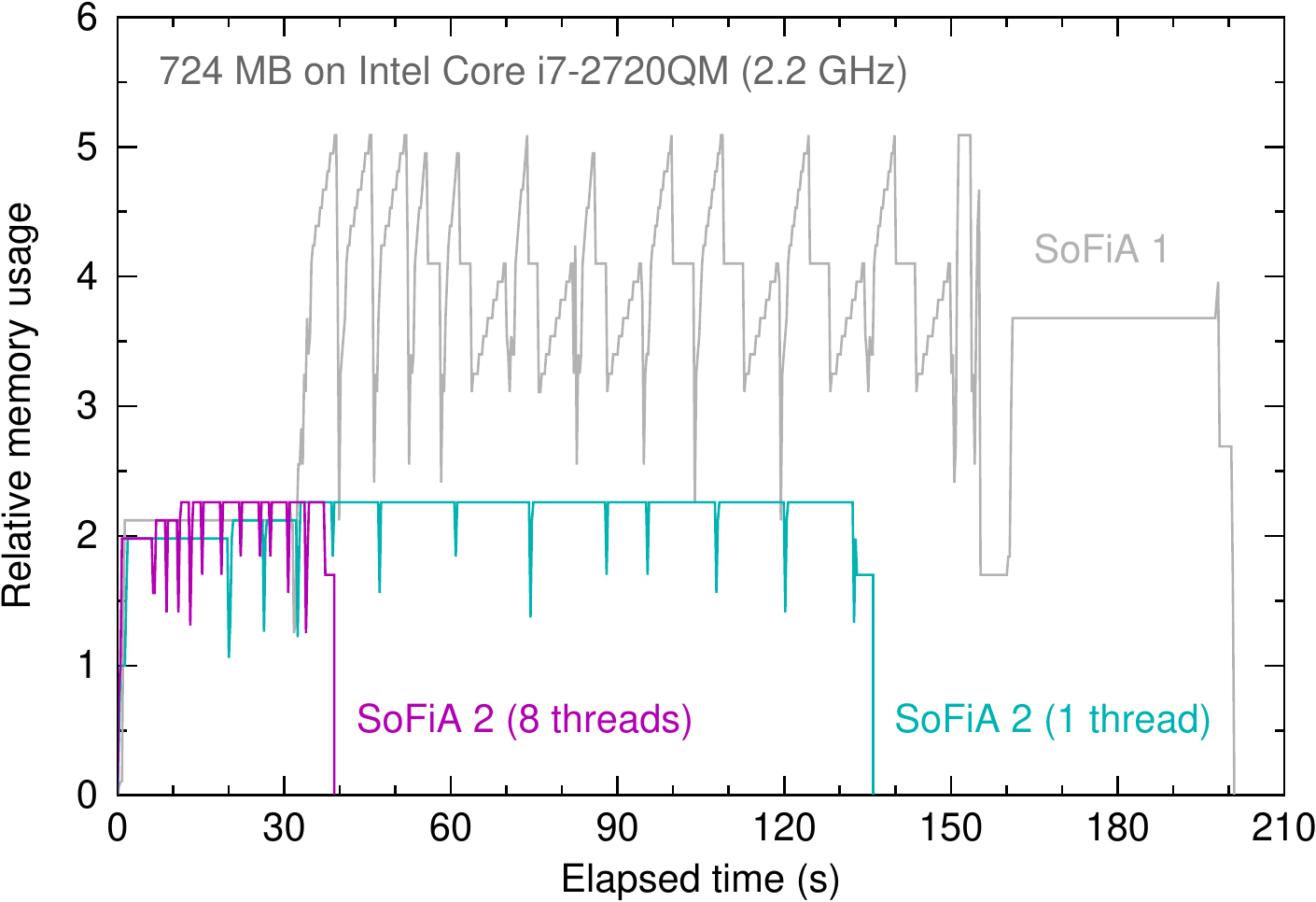} \hfill
		\includegraphics[height=5.9cm]{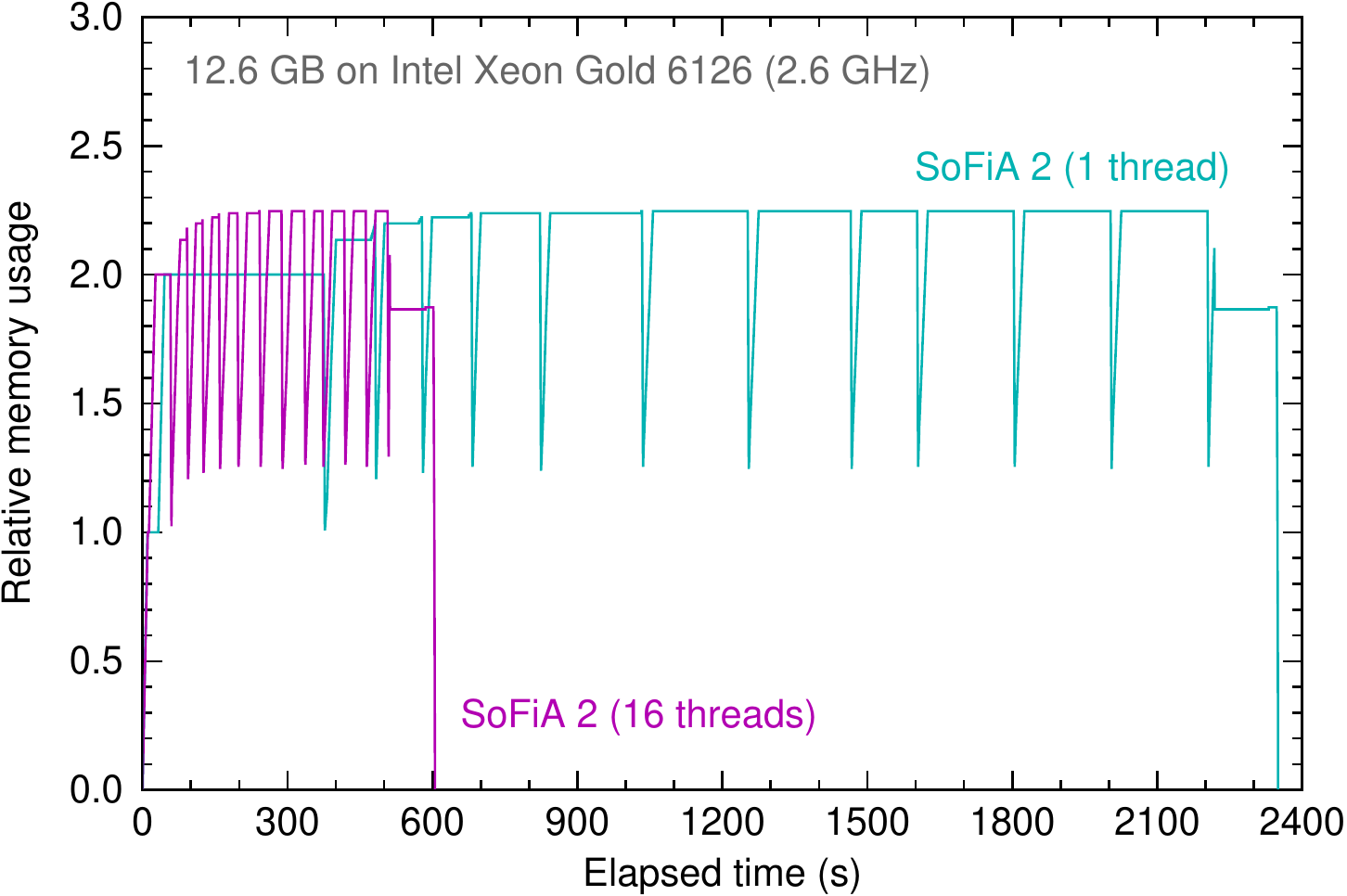}
		\caption{Left: Resident memory usage of \sofia{}~2 relative to the input data volume of 724~MB as a function of elapsed time for a single-threaded run (teal) and a multi-threaded run using 8~threads (magenta) of \sofia{}~2 in comparison to a similar run of \sofia{}~1 (grey) on a laptop computer equipped with Intel Core i7--2720QM CPUs clocked at 2.2~GHz. Right: Same, but for a run of \sofia{}~2 on a 12.6~GB data cube using 16~threads (magenta) versus 1~thread (teal) on one of the ICRAR Hyades cluster nodes equipped with Intel Xeon Gold 6126 CPUs clocked at 2.6~GHz.}
		\label{fig_performance}
	\end{figure*}

	\subsection{Reliability measurement}
	\label{sect_reliability}
	
	One of the most important features of \sofia{}~2 is its ability to statistically determine the reliability of each detection and use this information to automatically remove detections from the source catalogue that are deemed unreliable (e.g.\ \citealt{Dickinson2004,Yan2004,Kovac2009}). The underlying algorithm is described in detail in \citet{Serra2012a} and works by comparing the number density of detections with negative and positive total flux in a 3D parameter space made up of the peak flux density, the summed flux density and the mean flux density across the source.
	
	
	The reliability measurement is particularly powerful as it allows false positives to be automatically removed from the source catalogue by simply applying a reliability threshold. This allows \sofia{}~2 to be pushed deeper than other source finders by applying extreme detection thresholds of as low as 3 times the noise level without being overwhelmed by the large number of false positives that would otherwise result from such a choice.
	
	
	
	In addition to the reliability threshold, the user can also set a signal-to-noise threshold. All detections with an integrated signal-to-noise ratio below that threshold will be discarded as unreliable irrespective of their measured reliability. This feature relies on accurate beam information to be present in the data cube header and can be used to automatically remove detections that are too faint to be considered reliable and were assigned a high reliability value purely because of low number statistics in poorly populated regions of parameter space.
	
	Users are advised that the integrated signal-to-noise ratio of sources in the final source catalogue could potentially be lower than the threshold defined in the reliability module, as the reliability calculation is carried out on the noise-normalised data cube rather than the original cube. In addition, optional features such as mask dilation have the potential to alter the source mask and hence the measured signal-to-noise ratio of all detections.

	

	\subsection{Parameterisation and output}
	
	\sofia{}~2 will provide measurements of basic source parameters across the three-dimensional source mask. By default, these will be specified in the native pixel coordinates and flux units of the data cube, although the user has the option of manually enabling certain physical corrections, including conversion from pixel coordinates ($x$, $y$ and $z$) to proper world coordinates (e.g.\ celestial coordinates and frequency in the native units of the data cube) and division of spatially integrated flux parameters by the beam solid angle. \sofia{}~2 will extract the required data cube axis descriptors and beam information from the FITS file header, and it is the user's responsibility to ensure that the header information is correct and adequate. In addition, \sofia{}~2 implicitly assumes that the spectral channels of the data cube are uncorrelated. If this were not the case, e.g.\ due spectral smoothing, then the user would have to manually correct the relevant source parameters to account for the correlation of spectral channels.
	
	\sofia{}~2 will also derive basic statistical uncertainties for several fundamental parameters, including the centroid and integrated flux of the source. It should be emphasised that these are calculated assuming Gaussian error propagation and may not be representative of the true uncertainties which are often dominated by systematic rather than stochastic errors.
	
	\sofia{}~2 can save the final source catalogue in three different formats: plain-text format intended for basic visual inspection, VO-compliant XML format for use in Virtual Observatory tools, and SQL format for integration into a database. \sofia{}~2 can also provide more advanced data products for scientific and diagnostic purposes, including a copy of the source mask, global moment maps of all detections, diagnostic plots created by the reliability module and data products such as moment maps and integrated spectra for each individual detection.

	\begin{table}
		\centering
		\caption{Run time of \sofia{}~2 on a 12.6~GB test data cube with multi-threading enabled. The columns show the number of parallel threads used ($n_{\rm thr}$), the total run time ($t_{\rm run}$) excluding the time required to read the data cube, the total CPU time ($t_{\rm cpu}$), the speed-up factor relative to the run with 2~threads ($\eta$), and the resulting parallel fraction of the code ($f_{\rm p}$) using Eq.~\ref{eqn_amdahl}.}
		\label{tab_parfrac}
		\begin{tabular}{rrrrr}
			\hline
			$n_{\rm thr}$ & $t_{\rm run}$ & $t_{\rm cpu}$ & $\eta$ & $f_{\rm p}$ \\
			              &           (s) &           (s) &        &             \\
			\hline
			            2 &          1925 &          3291 &   1.00 &          -- \\
			            4 &          1155 &          3210 &   1.67 &        0.80 \\
			            8 &           772 &          3275 &   2.49 &        0.80 \\
			           16 &           582 &          3425 &   3.31 &        0.80 \\
			           32 &           474 &          3720 &   4.06 &        0.80 \\
			\hline
		\end{tabular}
	\end{table}

	\section{Speed and memory usage}
	\label{sect_speed_and_memory}
	
	Two of the most fundamental metrics of a source finding pipe\-line are the time it needs to process a given amount of data and the peak memory usage relative to the size of the data cube. We determined both metrics on a standard laptop computer with 8~GB of RAM and four Intel Core i7--2720QM CPUs clocked at 2.20~GHz with a total of 8~threads. The test was carried out on a data cube from \citet{Serra2012a} containing genuine noise from an \hi{} observation taken with the Westerbork Synthesis Radio Telescope and several injected galaxies obtained from the WHISP survey \citep{Swaters2002}. The data cube has a size of $360 \times 360$ spatial pixels and 1464~spectral channels, corresponding to a data volume of 724~MB. Two identical runs of \sofia{}~2 were carried out: one multi-threaded run using all 8~threads, and another single-threaded run with multi-threading disabled altogether. The test was carried out using version 2.2.1 of the software. In addition, we ran a comparable test on the same data cube using \sofia{}~1.3.3 to demonstrate the substantial improvement in speed and memory usage of \sofia{}~2.
	
	As the time required to read the data cube from disc into memory will depend on several external factors, the test was run with the data already cached in memory in order to measure just the processing time without the I/O contribution. We enabled local noise scaling across a spatial and spectral window size of 31~pixels/channels. We then employed the S+C finder using three spatial filters of 0, 5 and 10~pixels and four spectral filters of 0, 3, 7 and 15~channels. The flux detection threshold was set to 3.5 times the noise level in each smoothing iteration. In addition, we ran the linker, reliability filter and parameteriser before writing out the resulting catalogue to disc in all three output formats supported by \sofia{}~2. In addition, we created global moment maps as well as standard output products for each individual source (including sub-cubes, moment maps and spectra).
	
	The results of the performance test are presented in the left-hand panel of Fig.~\ref{fig_performance}. The resident memory usage of \sofia{}~2 peaks at just over 2.25 times the input data volume of 724~MB. This is exactly as expected for a 32-bit data cube, as the pipeline will need to hold at most two copies of the input cube plus an 8-bit source detection mask in memory at any one time. It also demonstrates the significant reduction in memory usage compared with \sofia{}~1 which requires memory equivalent to more than 5~times the input cube size.
	
	It took \sofia{}~2 about 136~s to complete the single-threaded run, corresponding to a processing rate of 19~GB per hour. With multi-threading enabled, the processing time reduces to just 39~s, corresponding to 65~GB per hour. Multi-threading therefore significantly reduces the overall processing time by a factor of 3.5 on a modest system with just 8~threads and despite the fact that not all of the algorithms in \sofia{}~2 can be, and have been, multi-threaded.
	
	In comparison, \sofia{}~1 required about 201~s to complete the test, which corresponds to 12.7~GB per hour and indicates that \sofia{}~2 is more than five times as fast as \sofia{}~1 even on a modest laptop computer and at the same time requires less than half the amount of memory.
	
	It should be noted that these performance measures were obtained with the input data cube already cached in memory, and additional time would be required to read the data from disc. The obtainable I/O speed will vary significantly as a function of several parameters, including the type of storage hardware used, the data access pattern (full cube versus sub-region) and any simultaneous I/O activity by other processes in the background. Hence, the performance measures obtained in our test are not expected to scale linearly with data cube size, and larger cubes may require disproportionately more time to process.
	
	\begin{figure*}
		\centering
		\includegraphics[width=0.67\linewidth]{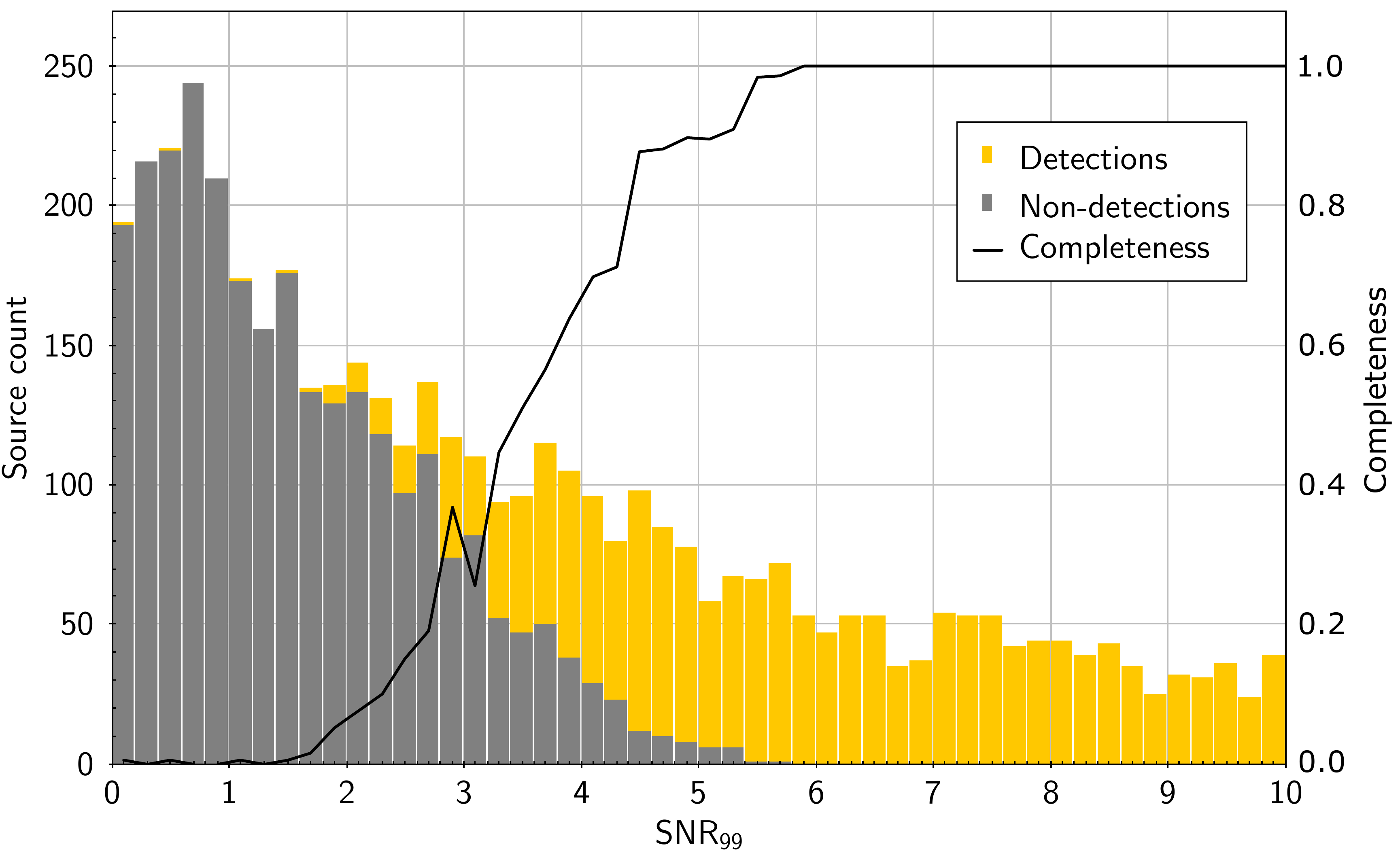}
		\caption{Stacked histogram of detected (yellow) and undetected (grey) mock galaxies and the resulting differential completeness (black curve) as a function of $\mathrm{SNR}_{99}$, demonstrating that \sofia{}~2 is able to achieve 90\% completeness at $\mathrm{SNR}_{99} \approx 5$.}
		\label{fig_completeness}
	\end{figure*}
	
	To assess the scalability of our performance test results, we repeated the speed and memory test with the same settings on a much larger data cube of 12.6~GB, this time using one of the nodes of the Hyades computing cluster at the International Centre for Radio Astronomy Research (ICRAR) in Perth, Western Australia, which is equipped with Intel Xeon Gold 6126 CPUs clocked at 2.6~GHz. The results are presented in the right-hand panel of Fig.~\ref{fig_performance}. When using just a single thread, \sofia{}~2 completed the run in about 39~minutes, corresponding to a data rate of 19~GB per hour. When utilising 16~threads simultaneously, the processing time decreased by a factor of 3.9 to about 10~minutes, implying a data rate of approximately 76~GB per hour. These values are in good agreement with those from our earlier test on a laptop computer, with a small improvement from the increase in the number of threads from 8 to 16. This suggests that the processing speed of \sofia{}~2 remains at a comparable level when scaling up the data volume by a factor of almost 20, and our performance measurements should therefore be robust and scalable.
	
	We can also use our run time measurements to estimate the parallel fraction of the code. In order to achieve this, we repeated the test run on the 12.6~GB data cube while varying the number of parallel threads available to \sofia{}~2. The resulting measurements are shown in Table~\ref{tab_parfrac}. As before, we exclude the initial time required to read in the data cube from the measured run time, $t_{\rm run}$. We then apply Amdahl's law \citep{Amdahl1967},
	\begin{equation}
		\frac{1}{\eta} = 1 - f_{\rm p} \left( 1 - \frac{1}{\Delta n_{\rm thr}} \right) , \label{eqn_amdahl}
	\end{equation}
	to measure the parallel fraction, $f_{\rm p}$, of the code, where $\eta$ is the speed-up and $\Delta n_{\rm thr}$ is the factor by which the number of threads has increased. With $f_{\rm p} \approx 0.8$, our measurements indicate that approximately 80~per cent of the code is running in parallel, which is an excellent outcome. This result also suggests that \sofia{}~2 should ideally utilise between 8 and 16~threads, beyond which any additional gains in speed become insignificant.
	
	Lastly, users should bear in mind that processing times, memory requirements and parallelisation efficiency will critically depend on the actual algorithms and parameter settings chosen by the user, and specific source finding runs could require more time and memory than the test run presented here. However, we deliberately activated settings that would be typical for a \wallaby{} source finding run, and the test result should therefore be fairly representative for the typical needs of extragalactic \hi{} source finding.

	\section{Completeness, reliability and parameterisation accuracy}
	\label{sect_performance}
	
	In order to assess the performance of \sofia{}~2 with respect to completeness, reliability and parameterisation accuracy, we created a mock data cube by injecting model galaxies into a data cube containing genuine ASKAP noise. The noise was obtained by extracting a subregion of $1501 \times 1501$ spatial pixels and $1501$ spectral channels from a \wallaby{} pre-pilot data cube of the Eridanus galaxy cluster, resulting in a file size of about $12.6~\unit{GB}$. The spatial pixel size of the cube is $6''$ (with a beam size of $\approx 30''$), and the spectral channel width is about $18.5~\unit{kHz}$ which corresponds to a velocity resolution of about $4~\unit{km \, s}^{-1}$ at redshift~0. The noise cube was extracted from the frequency range of $1323$ to $1351~\unit{MHz}$ to minimise the risk of contamination with genuine \hi{} emission. We then used the noise scaling algorithm of \sofia{}~2 to normalise the noise level to $1$, thus allowing us to conveniently specify all source parameters in units of signal-to-noise ratio.
	
	Next, we generated several thousand model galaxies using the \acronym{galmod} task of the \acronym{gipsy} data processing software \citep{Allen2011}. To ensure that the resulting galaxies would cover a wide range of different observational properties, we randomly varied several galaxy parameters within a meaningful range, including peak \hi{} column density ($10^{20}$\,--\,$10^{21}~\unit{cm}^{-2}$), rotation velocity ($30$\,--\,$220~\unit{km \, s}^{-1}$), exponential scale length on the sky ($4.5''$\,--\,$36''$), disc inclination ($0\degr{}$\,--\,$85\degr{}$) and position angle on the sky ($0\degr{}$\,--\,$360\degr{}$).
	
	The model galaxies were convolved with a $30''$ Gaussian beam in accordance with the expected restoring beam of the WALLABY data. $3200$ model galaxies were then injected into the noise cube on a regular grid in an attempt to fit as many model galaxies as possible. The flux density of each model galaxy was scaled by a constant factor to ensure that the integrated signal-to-noise ratios of most of the galaxies would fall into the range of about $0$ to $10$, as this is the most interesting range within which we would expect the completeness to increase from $0$ to $1$ (at $\mathrm{SNR} \approx 5$).
	
	In addition, we created a second version of the data cube by using the same set of galaxies, but this time using five times the original flux scaling factor to extend our sample into the higher signal-to-noise ratio range of up to about $50$ for the purpose of checking the source parameterisation accuracy of \sofia{}~2. The two SNR samples combined therefore provide us with an overall sample of $6400$ mock galaxies.
	
	Both model data cubes were then processed with \sofia{}~2 using the S+C finder with spatial kernels of $0$, $5$ and $10$ pixels and spectral kernels of $0$, $3$, $7$ and $15$ channels. In addition, we set the detection threshold to $3.5$ times the noise level, the linker radius to $2$ pixels/channels and the minimum source size to $8$ spatial pixels and $5$ spectral channels. Lastly, we enabled the reliability filter to automatically remove unreliable detections, setting a reliability threshold of $0.9$, a kernel scale factor of $0.4$ and a signal-to-noise threshold of $2.8$. 

	\subsection{Completeness and reliability}
	
	Among the most fundamental performance indicators of any source finding algorithm are its \emph{completeness} and \emph{reliability}. Completeness, $C$, is defined as the fraction of genuine sources being successfully detected by the source finder, hence
	\begin{equation}
		C = \frac{N_{\rm gen}}{N_{\rm tot}},
	\end{equation}
	where $N_{\rm gen}$ is the number of genuine sources detected, while $N_{\rm tot}$ is the total number of sources present. Likewise, reliability, $R$, is defined as the fraction of detected sources that are genuine, $N_{\rm gen}$, as opposed to false positives due to artefacts or noise, $N_{\rm false}$, hence
	\begin{equation}
		R = \frac{N_{\rm gen}}{N_{\rm gen} + N_{\rm false}}.
	\end{equation}
	Completeness and reliability are most meaningfully defined in differential form as a function of integrated signal-to-noise ratio (SNR), with both completeness and reliability expected to approach 100\% at high SNR, while gradually decreasing towards low SNR.
	
	Likewise, completeness and reliability of a source finding run will strongly vary with the parameter settings of the source finder, most notably the detection threshold. Using a lower detection threshold will increase the number of genuine detections at low SNR and hence completeness, while at the same time producing more false detections and thus decreasing reliability. Therefore, the main challenge with automated source finding is to establish optimal settings that strike a balance between acceptable levels of completeness and reliability.

	\subsubsection{Reliability}
	
	In order to obtain an estimate of the expected reliability of \sofia{}~2, we first ran the pipeline on the original noise cube \emph{without} injected mock galaxies, using exactly the same settings as for the two data cubes with galaxies injected. If \sofia{}~2 were fully reliable, this experiment should yield no detections, as the cube should only contain stochastic noise, albeit genuine noise produced by the ASKAP telescope and receiver system.
	
	Overall, \sofia{}~2 reported $12328$ detections at the $3.5 \sigma$ detection threshold chosen for this experiment. Of these, $6370$ have positive total flux, while $5958$ have negative flux. Virtually all of these detections are deemed unreliable by the reliability filter in \sofia{}~2, and only a single detection with positive flux remains after reliability filtering. Within just $5''$ of the measured sky position of that detection of $\alpha = 03^{\rm h}38^{\rm m}34^{\rm s}$ and $\delta = -22\degr{}46'08''$ there is a bright optical counterpart, LEDA~809162, which is classified as a galaxy in the NASA/IPAC Extragalactic Database.\footnote{\url{https://ned.ipac.caltech.edu/}} This suggests that the sole signal detected by \sofia{}~2 in the noise cube is a genuine high-redshift \hi{} emission line at $z = 0.066$ associated with LEDA~809162. Unfortunately, no optical redshift measurement is available for LEDA~809162 to unambiguously confirm that the \hi{} signal is genuine and indeed associated with that galaxy.
	
	With the sole detection being almost certainly a genuine galaxy at higher redshift, we can therefore be confident that no false positives were picked up by \sofia{}~2 after reliability filtering, indicating that the output catalogue produced by \sofia{}~2 is 100\% reliable at all SNR levels. This outcome highlights the excellent quality of ASKAP data as well as the power of the reliability filter implemented in \sofia{}~2 which appears to be capable of accurately discarding even a large number of false positives generated as a result of the low detection threshold of $3.5 \sigma$ applied in this experiment.
	
	It should be emphasised at this point that the reliability filter in \sofia{}~2 is based on the assumption that the image noise is symmetric about zero. The reliability of the source catalogue from \sofia{}~2 will therefore critically depend on how clean the underlying data are. In particular, effects such interference, residual continuum emission, \hi{} absorption, etc.\ are likely to reduce the effectiveness of the reliability filter, potentially reducing both the reliability and completeness of the resulting source catalogue. Later on in Section~\ref{sect_pilot} we will see an example of an actual ASKAP data set that is not as clean as the ASKAP noise used in our mock data set, resulting in a reduced effectiveness of the reliability filter and hence a somewhat lower reliability of the source catalogue.

	\begin{figure}
		\centering
		\includegraphics[width=\linewidth]{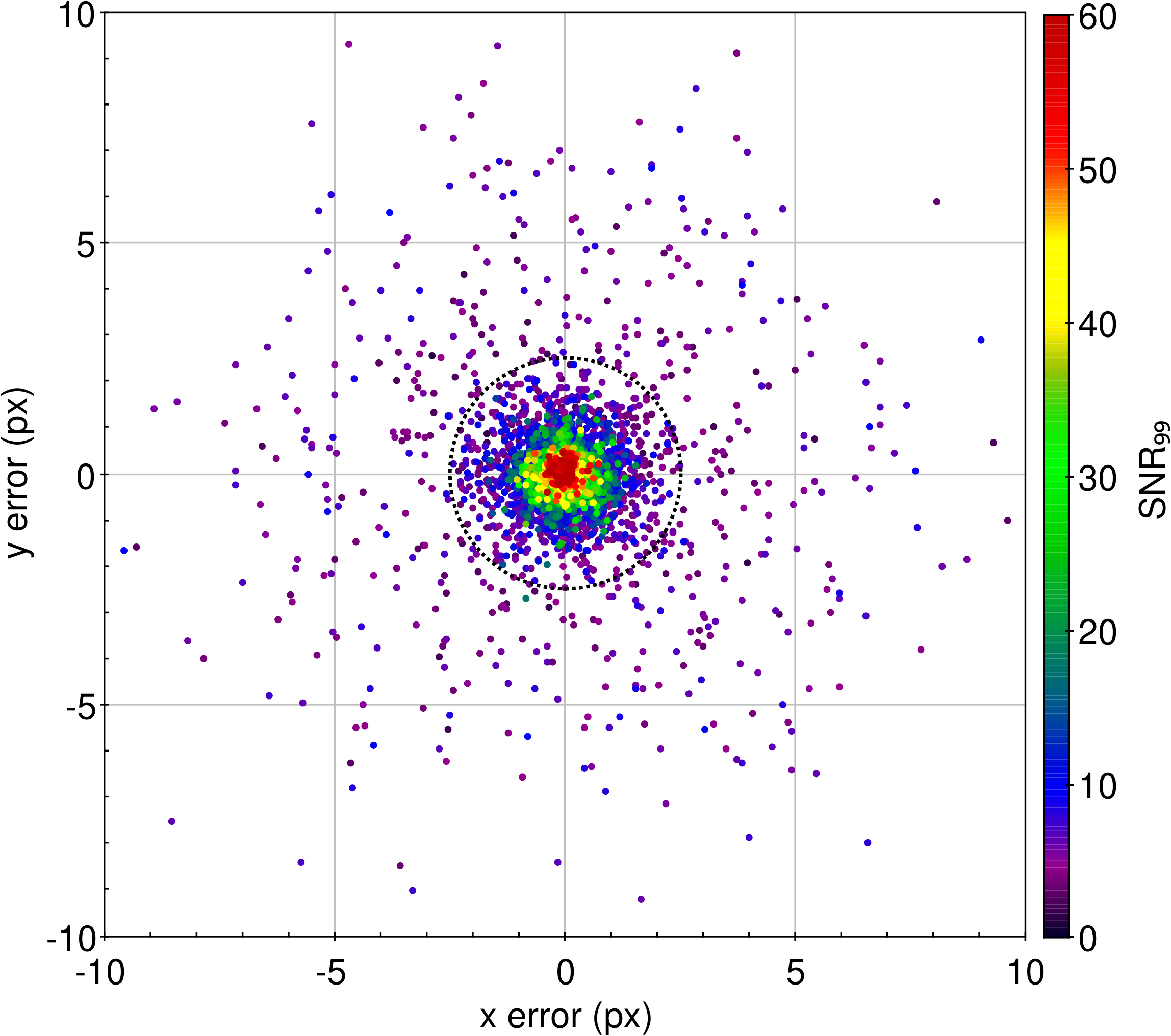}
		\caption{Position measurement errors colour-coded by $\mathrm{SNR}_{99}$ of each mock galaxy. The black, dotted circle marks the size of the 30-arcsec beam.}
		\label{fig_xy_err}
	\end{figure}

	\subsubsection{Completeness}
	
	In order to analyse the completeness of the source finding run as a function of signal-to-noise ratio, we will first need to establish a way of characterising the integrated signal-to-noise ratio (SNR) of a source. The integrated flux of a source is given by
	\begin{equation}
		F = \frac{\Delta \nu}{\Omega} \sum \limits_{i = 1}^{N} S_{i} \label{eqn_flux}
	\end{equation}
	where $\Delta \nu$ is the width of a frequency channel, $\Omega$ is the solid angle of the restoring beam in units of pixels and $S_{i}$ is the flux density value of pixel, $i$, with the summation carried out over all $N$ pixels considered to be part of the source. For a Gaussian beam, $4 \ln(2) \, \Omega = \uppi \, a \, b$, with $a$ and $b$ being the full width at half maximum of the major and minor axis of the beam in units of pixels. Note that we have made the explicit assumption that the spectral channels of the data cube are uncorrelated, which is certainly the case with our ASKAP data.
	
	In principle, we can use the Gaussian error propagation law to determine the statistical uncertainty of the flux measurement, thus
	\begin{equation}
		\sigma_{F} = \frac{\sqrt{N} \Delta \nu \, \sigma_{\rm rms}}{\sqrt{\Omega}} \label{eqn_flux_unc}
	\end{equation}
	where $\sigma_{\rm rms}$ is the original noise level per pixel and we have additionally accounted for the fact that spatial pixels are partially correlated due to the finite size of the beam, while we assume spectral channels to be entirely uncorrelated. The integrated signal-to-noise ratio is then simply given as $\mathrm{SNR} = F / \sigma_{F}$.
	
	While this method works well with detected sources, yielding the observed signal-to-noise ratio, $\mathrm{SNR_{obs}}$, it will be more challenging to apply to non-detections or mock galaxies. A practical solution would be to sum the pixels of the mock source in the order of decreasing flux density until a certain fraction of the total flux is reached. Using a cut-off of 99\% of the total flux, we have made use of this method here to define $\mathrm{SNR}_{99}$ for the purpose of calculating completeness. A more detailed description of the definition of $\mathrm{SNR}_{99}$ and its relation to the observed signal-to-noise ratio is given in Appendix~\ref{sect_snr}.
	
	The completeness resulting from our mock galaxy experiment is shown in Fig.~\ref{fig_completeness} as a function of $\mathrm{SNR}_{99}$. Apparently, \sofia{}~2 is capable of achieving 90\% completeness at $\mathrm{SNR}_{99} \approx 5$, with all galaxies above $\mathrm{SNR}_{99} = 6$ being picked up. Interestingly, \sofia{}~2 will still detect a few galaxies at very low signal-to-noise ratios in the range of 2--3 that must have been boosted by contributions from collocated noise peaks. Even at $\mathrm{SNR}_{99} = 3.5$ the completeness is still sitting at the 50\% level.
	
	The outcome of the mock galaxy experiment demonstrates that the S+C source finding algorithm of \sofia{}~2 in combination with its sophisticated reliability filter is capable of extracting a highly complete and reliable source catalogue from a large \hi{} data cube. As said before, the performance of \sofia{}~2 will stand and fall with the ability of the reliability filter to remove false detections. The cleaner the data, the better the algorithm is expected to perform, while the presence of artefacts will likely degrade the reliability and completeness of the source catalogue.
	
	Nevertheless, the experiment presented here was carried out using real noise from an actual spectral-line observation with ASKAP, demonstrating that optimal performance can be achieved on real data and that \sofia{}~2 is in principle ready to handle \hi{} data from the \wallaby{} survey.

	\subsection{Parameterisation accuracy}
	
	Another important performance metric to assess is the accuracy with which \sofia{}~2 is able to recover basic observational parameters of the mock galaxies, such as position, integrated flux or spectral profile width. As with completeness and reliability, this is expected to be a function of integrated signal-to-noise ratio, as measurements of fainter sources will suffer from larger statistical uncertainties and potentially be more susceptible to systematic errors such as the loss of emission from the faint outer regions of a source.

	\begin{figure}
		\centering
		\includegraphics[width=\linewidth]{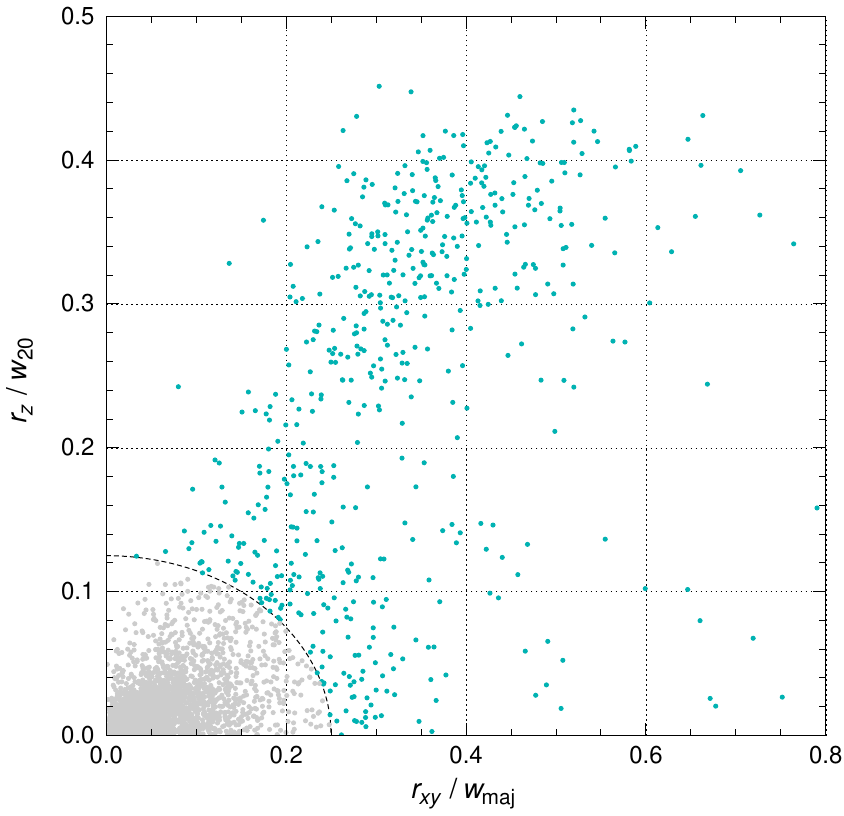}
		\caption{Ratio of spectral position error and spectral line width, $r_{z} / w_{20}$, plotted against the ratio of spatial position error and spatial major axis size, $r_{xy} / w_{\rm maj}$. The black, dashed curve marks the elliptical boundary of $r_{xy} < 0.25 \, w_{\rm maj}$ and $r_z < 0.125 \, w_{20}$ that we use to separate sources that have been left intact (grey data points) from sources that are likely to have been fragmented (teal data points) by \sofia{}~2.}
		\label{fig_fractured}
	\end{figure}

	\subsubsection{Source location}
	
	The position measurement errors from the mock galaxy experiment are presented in Fig.~\ref{fig_xy_err}. As can be seen, the positions produced by \sofia{}~2 are on average highly accurate with a mean of $0.0 \pm 1.6$ pixels in both $x$ and $y$, the standard deviation corresponding to about one third of the size of the beam. As expected, the position accuracy is a function of integrated signal-to-noise ratio of the source, with a standard deviation of about $0.5$~pixels for brighter galaxies of $\mathrm{SNR}_{99} > 10$ as compared to about $2.1$~pixels for fainter galaxies of $\mathrm{SNR}_{99} < 10$.

	While most mock galaxies are located within the size of the beam, Fig.~\ref{fig_xy_err} does reveal a faint halo of points beyond the beam size. This halo is likely to have been caused by sources that were either just partially detected or fragmented into multiple components by \sofia{}~2. This is known to occasionally occur in the case of very faint, edge-on galaxies where the two halves of the galaxy are picked up as separate detections.
	
	The situation is very similar for the location accuracy along the frequency axis, where the mean error is $0.0 \pm 8.6$~channels. Again, the standard deviation is affected by outliers due to partially detected or fragmented galaxies at low signal-to-noise ratio. When only considering galaxies with $\mathrm{SNR}_{99} > 10$, the standard deviation drastically reduces to just 0.9~channels, demonstrating the excellent recovery of frequency centroids by \sofia{}~2.

	\begin{figure*}
		\includegraphics[width=\linewidth]{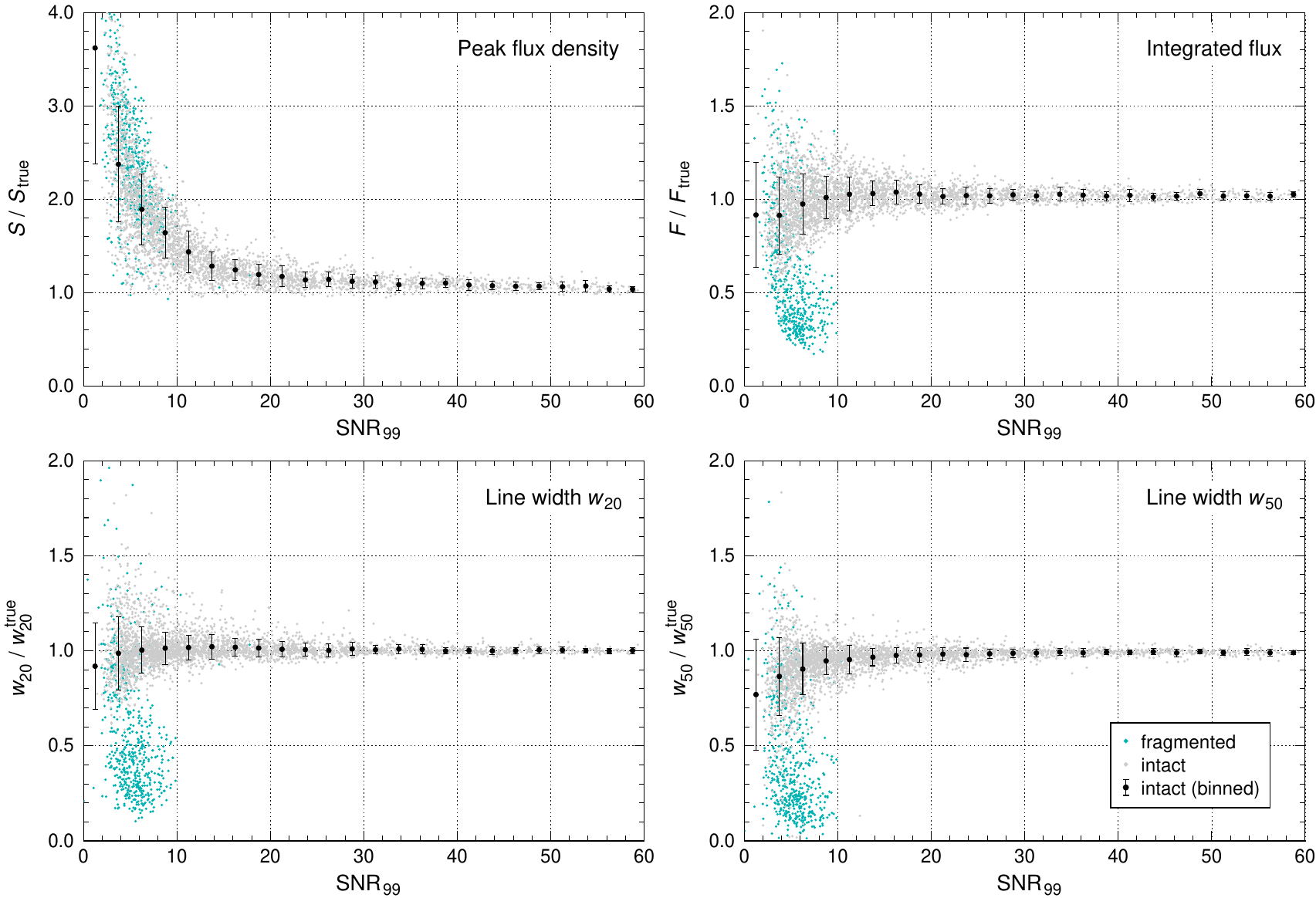}
		\caption{Ratio of measured versus true source parameter as a function of $\mathrm{SNR}_{99}$ for peak flux density (upper-left), integrated flux (upper-right), $w_{20}$ line width (lower-left) and $w_{50}$ line width (lower-right). Intact sources are shown in grey, fragmented ones in teal. The black data points and error bars show the mean and standard deviation of the grey data points in intervals of 2.5.}
		\label{fig_param}
	\end{figure*}

	\begin{figure*}
		\includegraphics[width=\linewidth]{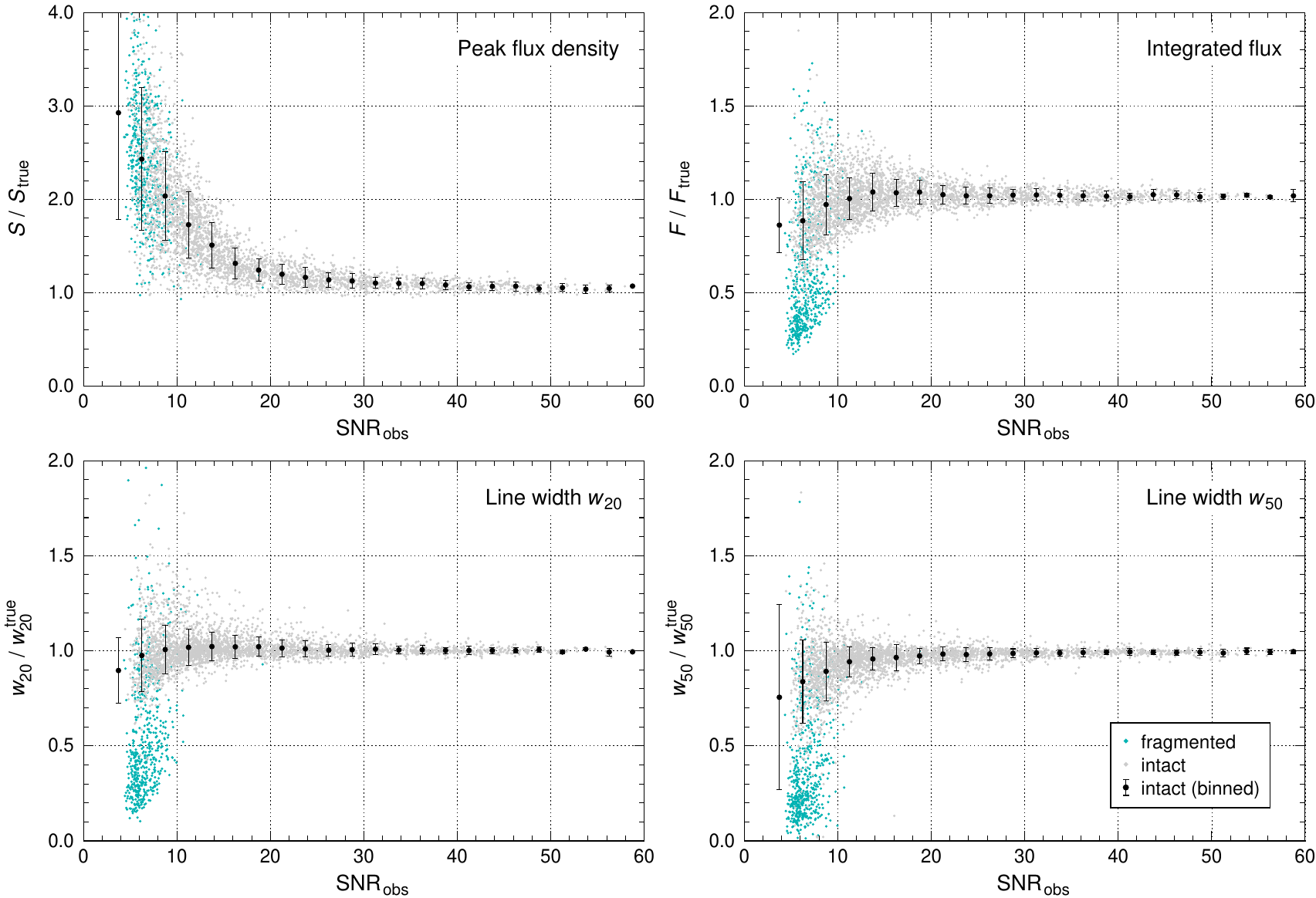}
		\caption{Same as Fig.~\ref{fig_param}, but as a function of observed signal-to-noise ratio, $\mathrm{SNR_{obs}}$.}
		\label{fig_param2}
	\end{figure*}

	\subsubsection{Fragmented sources}
	
	To further investigate the issue of fragmented detections, we plot in Fig.~\ref{fig_fractured} the ratio of spectral position error and spectral line width, $r_{z} / w_{20}$, against the ratio of spatial position error and spatial major axis size, $r_{xy} / w_{\rm maj}$, for all mock galaxies detected by \sofia{}~2. There is a strong concentration of sources near the origin that are likely to have been left intact by the source finder, as their position errors are very small compared to their spatial and spectral extent.
	
	In contrast, the population of fragmented sources is clearly visible as the extended halo of detections occupying the positional error range of roughly 20 to 50 per cent of the source extent. In order to separate between intact and fragmented sources, we apply a conservative threshold in the form of an elliptical radius of $r_{xy} < 0.25 \, w_{\rm maj}$ and $r_z < 0.125 \, w_{20}$, as marked by the dashed curve in Fig.~\ref{fig_fractured}. We consider all detections within that threshold to be intact (grey data points), while most, if not all, of the fragmented sources are expected to be located outside that threshold (teal data points). We will maintain this cut and the associated colour scheme throughout the remainder of this section to be able to separate between intact and fragmented sources in our analysis.
	
	The fraction of sources that have potentially been fragmented by \sofia{}~2 is a strong function of signal-to-noise ratio. At $\mathrm{SNR}_{99} = 5$ approximately 34\% of the detections are located outside the relative position error threshold considered here, whereas that fraction effectively drops to zero beyond $\mathrm{SNR}_{99} = 10$, suggesting that sources near the detection threshold are most susceptible to being fragmented, as would be expected.
	
	Likewise, the fraction of fragmented sources varies strongly with spectral line width. While less than 10\% of galaxies with narrow lines of $w_{20} \lesssim 150~\mathrm{km \, s}^{-1}$ are getting fragmented in our specific experiment, that fraction gradually increases to about 40\% for galaxies with a broad spectral profile of $w_{20} \gtrsim 400~\mathrm{km \, s}^{-1}$. Again, this is expected at low SNR levels, as edge-on galaxies with large rotation velocities are most susceptible to being fragmented by the source finder due to their broad double-horn profiles.
	
	As fragmented sources have the potential to significantly alter the results of certain scientific studies, e.g.\ measurements of the \hi{} mass function, they would need to be handled separately. Identification of partially detected or broken-up sources could potentially be achieved through comparison with optical catalogues or identification of close pairs of detections with similar properties.

	\subsubsection{Peak flux density}
	
	The peak flux density of a source reported by \sofia{}~2 is simply the highest flux density value encountered within the 3D source mask (not the peak flux density of the integrated spectrum). As such, the peak flux density will be strongly affected by the presence of noise which should result in a significant positive bias. This is illustrated in the upper-left panel of Fig.~\ref{fig_param} and~\ref{fig_param2} where the ratio of measured versus true peak flux density of the mock galaxies is plotted as a function of $\mathrm{SNR}_{99}$ and $\mathrm{SNR_{obs}}$, respectively.\footnote{Parameterisation accuracy is presented here as the ratio of measured versus true value, which is equal to the relative error plus one and better accounts for the multiplicative nature of the flux measurement process in radio astronomy.} The significant bias, in particular below $\mathrm{SNR}_{99} \approx 10$, means that the peak flux density is not a particularly meaningful parameter beyond the purpose of basic sanity checking, and its use in any form of scientific analysis is therefore discouraged.

	\subsubsection{Integrated flux}
	
	One of the most fundamental observational parameters to be extracted from any source finding effort is the integrated flux of a source, as it is required for determining the \hi{} mass of a galaxy. \sofia{}~2 measures the integrated flux, $F$, by summing the flux density values, $S_{i}$, of all $N$ pixels contained in the source mask, additionally multiplying by the spectral channel width, $\Delta \nu$, and dividing by the beam solid angle, $\Omega$, if explicitly requested to do so by the user (Eq.~\ref{eqn_flux}).
	
	In addition, \sofia{}~2 calculates the statistical uncertainty of the integrated flux measurement by assuming Gaussian error propagation and correcting for the fact that spatial pixels will be correlated due to the finite beam size (Eq.~\ref{eqn_flux_unc}). \sofia{}~2 measures the local noise level in the vicinity of each detection for this purpose.
	
	The flux measurement errors resulting from the mock galaxy experiment are presented in the upper-right panel of Fig.~\ref{fig_param} and~\ref{fig_param2} as a function of $\mathrm{SNR}_{99}$ and $\mathrm{SNR_{obs}}$, respectively. For galaxies not broken up into multiple detections (grey and black data points) the integrated flux is accurately recovered across the entire range of signal-to-noise ratios, with the statistical uncertainty increasing towards lower $\mathrm{SNR}_{99}$, as expected. A closer inspection of Fig.~\ref{fig_param} reveals that the mean relative flux measurement error is not exactly centred on zero. Instead, the flux measurement from \sofia{}~2 is on average about 2--3\% higher across most of the SNR range. This effect is particularly noticeable at higher $\mathrm{SNR}_{99}$ where the statistical uncertainties are smaller.
	
	Our investigation of this flux measurement bias revealed that it is likely caused by a particular aspect of the S+C finder whereby at the beginning of each iteration all pixels already detected in previous iterations are set to $\pm n$ times the noise level (preserving the original sign of the flux value) in the data cube before smoothing to prevent the smoothing operation from smearing out the emission beyond the boundaries of the source. A side effect of this approach appears to be that the resulting source mask tends to be grown in directions where a predominantly positive contribution from the noise can be expected, thus resulting in a net positive flux bias.
	
	The effects appears to depend directly on the replacement value chosen. The default value of $n = 2$ used here will result in the observed 2--3\% positive flux bias, while we have found the bias to disappear entirely when slightly reducing the value to $1.5$, albeit at the cost of a somewhat stronger negative flux bias at the faint end of the SNR range. If high flux accuracy of better than 3\% is required, a statistical correction would need to be applied to remove any bias.
	
\begin{figure}
		\includegraphics[width=\linewidth]{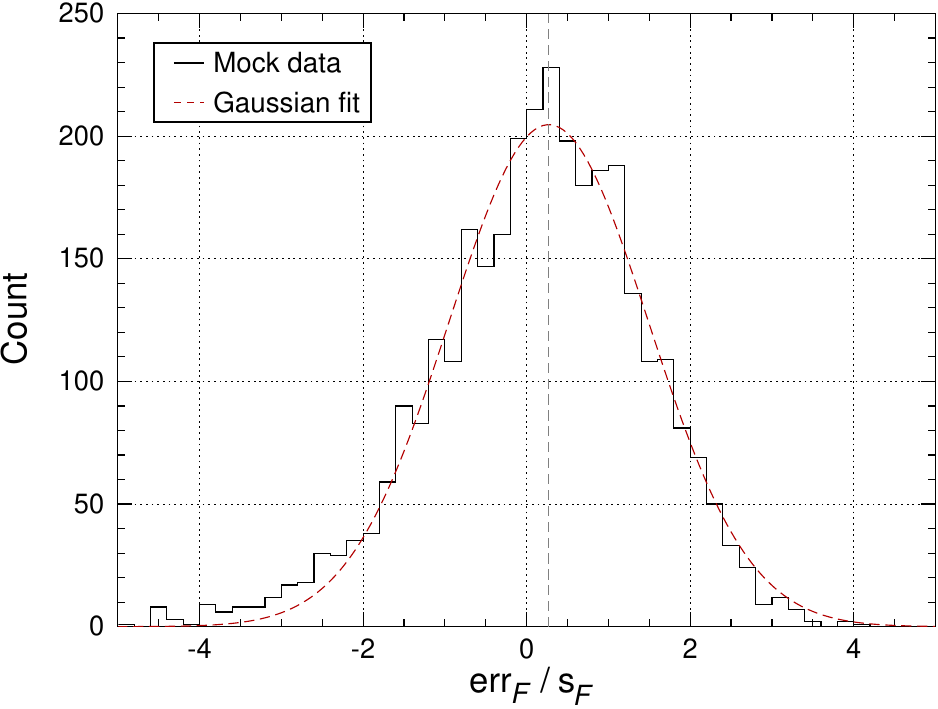}
		\caption{Histogram of the true integrated flux error, $\mathrm{err}_{F} = F_{\rm obs} - F_{\rm model}$, divided by the integrated flux uncertainty, $\sigma_{F}$, derived by \sofia{}~2 for the mock galaxies. The dashed, red line shows the result of a Gaussian fit to the histogram which should ideally be centred at $0$ with a standard deviation of $1$. The vertical, dashed, grey line marks the actual peak position of $0.27$, which is very close to the expected value.}
		\label{fig_fint_err}
	\end{figure}

	In Fig.~\ref{fig_fint_err} we show a histogram of the flux measurement error divided by the flux measurement uncertainty from \sofia{}~2. Under perfect conditions, $\mathrm{err}_{F} / \sigma_{F}$ should have a Gaussian distribution with a centroid of $0$ and a standard deviation of $1$. A Gaussian fit to the histogram yields a centroid of $0.27 \pm 0.02$ and a standard deviation of $1.22 \pm 0.02$, which is very close to the expected values and indicates that the flux uncertainties reported by \sofia{}~2 are accurate. The small offset and slight broadening of the distribution are not unexpected and likely reflect the minor flux bias and its variation with signal-to-noise ratio.

	\subsubsection{Spectral line width}
	
	Another fundamental observational parameter to be extracted by any source finder is the width of the integrated spectral profile of a galaxy which forms the basis for determining the rotation velocity. \sofia{}~2 determines the integrated spectral profile by summing over the spatial pixels contained in the source mask in each spectral channel. The $w_{20}$ and $w_{50}$ line widths are then measured by moving inwards from both ends of the spectrum until the signal exceeds 20\% or 50\% of the peak flux density in the spectrum, respectively. For improved accuracy, \sofia{}~2 will linearly interpolate across the two channels in between which the signal exceeds the threshold for the first time.
	
	As the line width measurement depends on the peak flux density of the spectrum, it is systematically affected by the presence of noise which will result in an overestimation of the peak flux density and thus an underestimation of the line width. This particularly affects the $w_{50}$ measurement and is one of several reasons for why $w_{50}$ may not provide an accurate measurement of the line width and rotation velocity of a galaxy at low SNR (see also \citealt{Tully1985} and \citealt{Ho2007}). The measured $w_{50}$ errors of the mock galaxies are plotted in the lower-right panel of Fig.~\ref{fig_param} and~\ref{fig_param2} as a function of $\mathrm{SNR}_{99}$ and $\mathrm{SNR_{obs}}$, respectively. The significant negative bias, particularly at $\mathrm{SNR}_{99} \lesssim 10$, is clearly visible, and we therefore discourage the use of $w_{50}$ when dealing with objects of low signal-to-noise ratio.
	
	The $w_{20}$ measurement errors resulting from the mock galaxy experiment are presented in the lower-left panel of Fig.~\ref{fig_param} and~\ref{fig_param2} as a function of $\mathrm{SNR}_{99}$ and $\mathrm{SNR_{obs}}$, respectively. Unlike $w_{50}$, the $w_{20}$ measurement from \sofia{}~2 is highly accurate across the entire range down to $\mathrm{SNR}_{99} \approx 5$, again with the exception of those galaxies that got fragmented by the source finder. The measurement of $w_{20}$ is even more accurate than the integrated flux measurement, and there is no sign of any significant bias. As expected, the statistical uncertainty increases towards the faint end.
	
	One of the main reasons for why the measurement of $w_{20}$ by \sofia{}~2 is so accurate is the fact that the integrated spectrum is created across the three-dimensional source mask, the shape of which can be expected to closely match the morphology of the source. This has the effect of maximising the signal-to-noise ratio and suppressing noise near the edges of the spectrum, where only few spatial pixels are expected to contribute. As a consequence, the $w_{20}$ measurement is not quite as much affected by noise as in classical studies where the spectrum is usually integrated across a rectangular mask and thus much noisier near the edges of the profile.

	\begin{figure*}
		\centering
		\includegraphics[width=\linewidth]{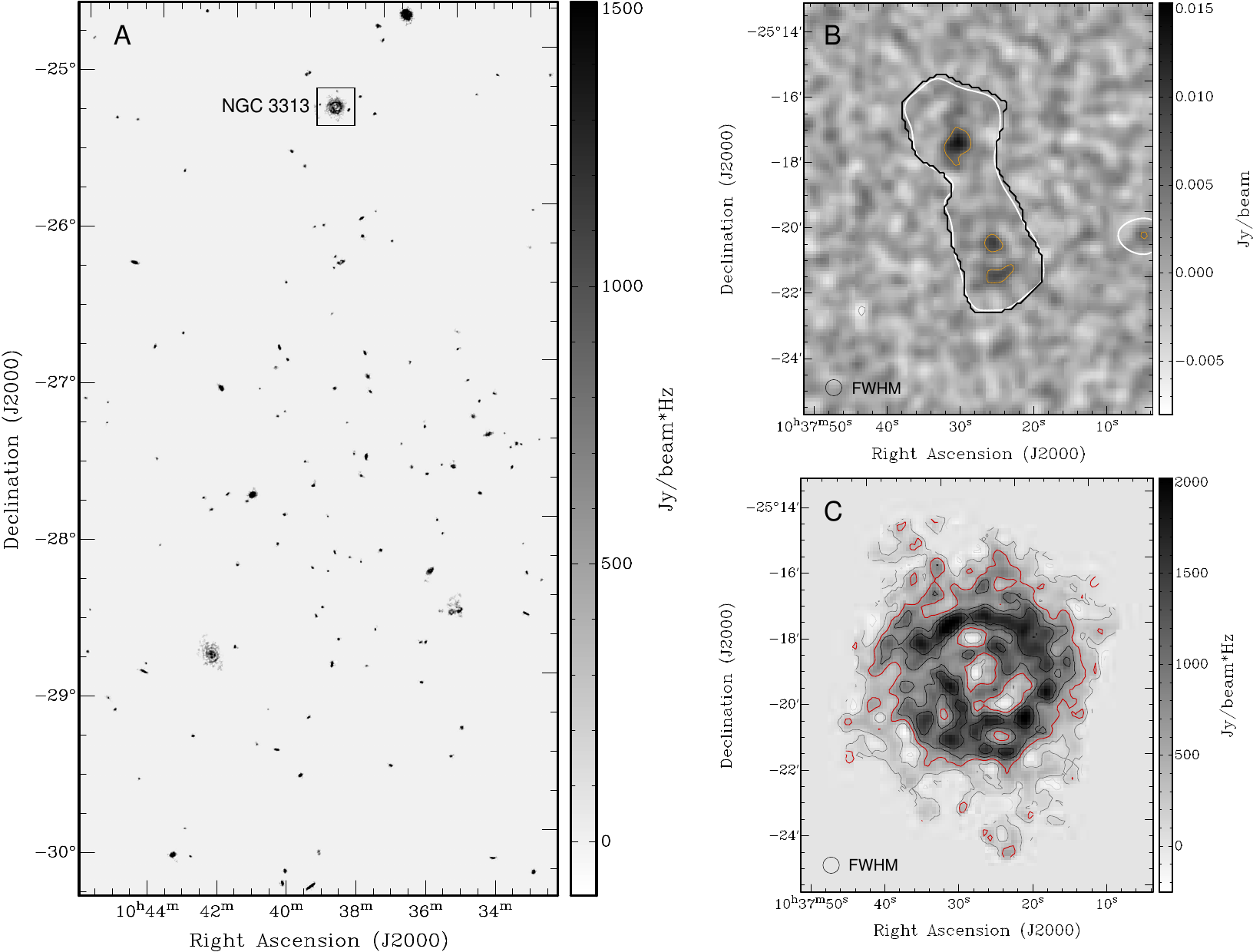}
		\caption{Panel~A: Moment-0 map from a \sofia{}~2 run on a $3\fdg{}0 \times 5\fdg{}5$ section of the \wallaby{} pilot survey field in the direction of the Hydra galaxy cluster. Almost 150 individual galaxies were detected by \sofia{}~2. Note that only pixels contained within the source mask are included in the moment calculation, and all other pixels are set to zero. Panel~B: An individual channel map from the original data cube in the direction of the galaxy NGC~3313 (see black box in panel~A). The bold, black contour shows the outline of the source mask produced by \sofia{}~2. The orange and grey contours correspond to $\pm 6~\mathrm{mJy}$ ($\approx \pm 3 \sigma$) in the original data cube. The bold, white contour corresponds to $2~\mathrm{mJy}$ ($\approx 3 \sigma$) after smoothing spatially with a Gaussian kernel of 10~pixels FWHM and spectrally with a boxcar filter of 15~channels to mimic the maximum smoothing scales applied by \sofia{}~2. Panel~C: The resulting moment~0 map of NGC~3313 integrated across the full 3-D source mask. The signal-to-noise ratio per pixel is shown as contours at levels of $-1$ (light-grey), $1$ (dark-grey), $3$ (bold, red), $5$ and $7$ (both black). Note that the additional white contour near the right-hand edge of panel~B is a satellite galaxy of NGC~3313 that was also detected by \sofia{}~2, but is not included in panel~C, as it is a separate source with its own mask. Also see Figure~14 in \citet{Koribalski2020} for a more detailed view of this data cube.}
		\label{fig_hydra}
	\end{figure*}

	\section{\wallaby{} pilot data}
	\label{sect_pilot}
	
	To demonstrate the fitness of \sofia{}~2 for full-scale \wallaby{} source finding, we ran a single-threaded version of the pipeline on the first \wallaby{} pilot survey data in the direction of the Hydra galaxy cluster using the parallel \mbox{\sofia{}-X} framework introduced in Section~\ref{sect_parallel}. Approximately 85~GB of data in a field of about $3\fdg{}0 \times 5\fdg{}5$ (equivalent to $1801 \times 3601$ pixels) on the sky out to a redshift of $z \approx 0.05$ (equivalent to 3517 frequency channels) were processed. The task was split across four nodes of the Hyades computing cluster at ICRAR. The output catalogues and images were directly written into a dedicated source database, and duplicate detections in the 300-pixel overlap bands between the four sub-regions were automatically identified and merged.
	
	Overall, \sofia{}~2 completed the source finding run in approximately 2~hours (without multi-threading), producing 197 unique detections after the automatic removal of duplicates. As a result of artefacts in the image data produced by the ASKAP data reduction pipeline \citep{Guzman2019}, some of these detections were found to be unlikely to be genuine astronomical sources and manually removed from the catalogue and output images. 148~detections remained after this process, and the moment-0 image resulting from the test run is presented in panel~A of Fig.~\ref{fig_hydra}. The faintest sources in this image have an \emph{observed} integrated signal-to-noise ratio of about~4. These sources are so faint that they are actually no longer visible to the naked eye in an individual channel map of the data cube, as their signal remains well below the noise level.
	
	To illustrate this effect, we show in panel~B of Fig.~\ref{fig_hydra} an individual channel map from the original data cube in the direction of the galaxy NGC~3313. The source mask produced by \sofia{}~2 in this frequency channel (bold, black contour) extends well beyond the brightest knots in the \hi{} disc of NGC~3313. The fainter regions of the disc are not visible at the original resolution of the data, as they are sitting below the noise level. Yet, thanks to its spatial and spectral smoothing, the S+C algorithm implemented in \sofia{}~2 is capable of detecting the full extent of the disc within the limits set by the sensitivity of the data. This is illustrated by the white contour in panel~B which corresponds to approximately three times the noise level of the data cube after spatial and spectral smoothing with the largest kernels that were actually applied by the S+C finder. The \sofia{}~2 source mask nicely traces the full extent of the disc after smoothing, as is also shown by the resulting moment~0 map of NGC~3313 across the full source mask in panel~C.
	
	The outcome demonstrates that \sofia{}~2 is capable of processing real \wallaby{} data in a short amount of time. In fact, with multi-threading enabled, the same run would have been completed in under 30~minutes if 16~CPU threads had been utilised simultaneously. Such processing times are negligible compared to the 16~hours it took to acquire the data at the telescope, let alone the additional substantial amount of time that was needed to calibrate and image the raw visibility data. The false positives picked up by \sofia{}~2 are the result of artefacts caused by insufficient flagging of the visibility data and inadequate continuum subtraction by the ASKAP data reduction pipeline. Efforts to improve the flagging and continuum subtraction algorithms are under way and expected to yield much cleaner data cubes that should get us significantly closer to the 100\% reliability achieved in our mock galaxy experiment. The mock data were based on real ASKAP noise from \wallaby{} pre-pilot observations of the Eridanus galaxy cluster which were less affected by such artefacts.
	
	We would also like to emphasise that for the Hydra data we deliberately pushed \sofia{}~2 to its limits by adopting a significantly lower reliability threshold (0.6 vs.\ 0.9) and smaller reliability kernel scale factor (0.3 vs.\ 0.4) than for the mock data test presented in Section~\ref{sect_performance}. This is expected to improve the completeness of the catalogue as compared to the mock data experiment, but at the expense of a reduction in reliability. The lower reliability obtained for the Hydra data is therefore not in tension with the mock data experiment, but rather a deliberate choice in our attempt to push for even greater completeness by challenging the limits of \sofia{}'s algorithms. If we had adopted the more conservative and robust settings of the mock data experiment, as will likely be the case for the full \wallaby{} survey, we would have obtained completeness and reliability values more in line with those presented in Section~\ref{sect_performance}.
	
	Lastly, the test run also demonstrates the ability of the parallel \mbox{\sofia{}-X} framework to split large-scale \wallaby{} data sets across multiple computing nodes and automatically merge the resulting catalogues and data products without any conflict. We are therefore confident at this stage that \sofia{}~2 fulfils all of the requirements for full-scale and real-time source finding on the large data volumes expected to be produced by the \wallaby{} survey in the near future.

	\section{Summary and conclusions}
	\label{sect_summary}
	
	\sofia{}~2 is a new extragalactic \hi{} source finding pipeline that is significantly faster and more memory-efficient compared to its predecessor, \sofia{}~1, thanks to the use of the C programming language in combination with multi-threading of time-critical algorithms. A substantial reduction in the number of third-party dependencies also means that \sofia{}~2 is also much easier to install than its predecessor. In addition, we have developed a parallel framework called \mbox{\sofia{}-X} that allows the processing of large data cubes to be split across multiple computing nodes in order to reduce processing times even further.
	
	Our performance tests of \sofia{}~2 on both mock data and genuine ASKAP data have demonstrated that the pipeline is capable of processing the large data volumes expected to be produced by the \wallaby{} survey in much better than real time when run in parallel on multiple computing nodes and with multi-threading enabled. In our tests we have been able to achieve processing rates of the order of 76~GB per hour using 16~threads on a single node of the Hyades cluster at ICRAR. This should in principal enable us to process each 800~GB \wallaby{} data cube in less than one hour if a sufficiently large machine with about 20~nodes and 16~threads per node could be utilised. Even if the number of available nodes and threads were lower, we would still be able to process the entire cube in much less than real-time (16~h observing time plus calibration and imaging), demonstrating that \sofia{}~2 is sufficiently fast at this stage to handle full-scale \wallaby{} survey data. In addition to this substantial increase in speed, \sofia{}~2 also occupies significantly less memory than its predecessor, \sofia{}~1, thus allowing much larger volumes of data to be loaded and processed simultaneously on a single node.
	
	Tests on mock galaxies injected into genuine ASKAP \hi{} data have shown that \sofia{}~2 is in principle capable of achieving close to 100\% reliability thanks to its built-in reliability measurement and filtering module, while still maintaining about 90\% completeness at an integrated signal-to-noise level of~5 and 100\% completeness above $\mathrm{SNR}_{99} \approx 6$. This result is encouraging and suggests that \wallaby{} will be able to achieve its anticipated detection rate and science goals. Users should note that their data will need to be relatively clean and free from major artefacts for the reliability filter to be fully effective.
	
	The accuracy of basic observational source parameters derived by \sofia{}~2 is generally high, and in particular spatial and spectral centroids, integrated fluxes and $w_{20}$ line widths are accurately recovered, with just minor biases of no more than a few percent across the full range of signal-to-noise ratios considered in our test. Recovery of the peak flux density and $w_{50}$ line width is more strongly affected by the presence of noise in the data and hence subject to significant biases, in particular at lower signal-to-noise ratios; their use in any kind of scientific analysis is therefore not recommended.
	
	There is a general risk of faint sources of $\mathrm{SNR}_{99} < 10$ getting broken up into multiple detections or being only partially detected by \sofia{}~2. This is particularly the case for faint, edge-on galaxies where the \hi{} emission is concentrated in the two `horns' of the spectrum, with very little emission in between. It should be noted that the fragmentation of sources is a direct consequence of the stochastic nature of the noise in combination with the low signal-to-noise ratio of some of the objects picked up by \sofia{}~2 and therefore in principle unavoidable.
	
	Mitigation strategies should be adopted to deal with both the issue of source fragmentation as well as parameterisation bias, as they would otherwise have the potential to corrupt any scientific analysis based on the source finding output. Possible strategies for identifying fragmented sources could involve the identification of close pairs of detections with otherwise similar parameters, or a comparison with optical position and redshift measurements to identify suspicious offsets. Alternatively, a signal-to-noise cut of about 10 can be used to obtain a clean catalogue that is largely unaffected by source fragmentation.
	
	Likewise, if extremely high parameterisation accuracy of better than a few per cent is required, a full-scale bias assessment using mock sources will need to be carried out similar to the mock data experiment presented here. This would also help with the establishment of completeness and reliability limits. Another possibility of reducing parameterisation biases would be the use of external software and algorithms for the parameterisation of sources detected by \sofia{}~2. As an example, fitting a Gaussian function or the Busy Function \citep{Westmeier2014} to the integrated spectrum provided by \sofia{}~2 could help in obtaining a more accurate measurement of the $w_{50}$ line width.
	
	As \sofia{}~2 is likely to be useful for other large-scale \hi{} surveys and potentially other spectral-line data, including observations of the CO line and other molecular transitions, the software has been made freely available to the entire community under an open-source licence and can be downloaded from GitHub. Feedback and bug reports from users of \sofia{}~2, either via e-mail or through the official GitHub issue tracking system, are encouraged and will help us to further improve the software and make it as useful and robust as possible for a wide range of spectral-line data.
	
	Lastly, we would like to emphasise again that all of the performance measures presented in this paper, including processing speed, memory usage, completeness, reliability, parameterisation accuracy and source fragmentation fraction, are specific to the data cubes and parameter settings used here and should not be relied upon. Other data sets are likely to require different settings and will almost certainly produce deviating performance metrics. Potential users of \sofia{}~2 are therefore advised to perform their own checks using mock data that more closely reflect the characteristics of their specific observational data.

	\section*{Acknowledgements}
	
	We wish to thank the referee for not only providing us with a constructive and helpful report, but also taking the time to install and test \sofia{}~2. This effort has led to several improvements to both the software and its documentation.
	
	The Australian SKA Pathfinder is part of the Australia Telescope National Facility which is managed by CSIRO. Operation of ASKAP is funded by the Australian Government with support from the National Collaborative Research Infrastructure Strategy. ASKAP uses the resources of the Pawsey Supercomputing Centre. Establishment of ASKAP, the Murchison Radio-astronomy Observatory and the Pawsey Supercomputing Centre are initiatives of the Australian Government, with support from the Government of Western Australia and the Science and Industry Endowment Fund. We acknowledge the Wajarri Yamatji people as the traditional owners of the Observatory site.
	
	The authors acknowledge support from the Astro\-nomy Data And Computing Services (ADACS) Software Support Programme and the Australian SKA Regional Centre (AusSRC) Design Study Programme.
	
	Parts of this research were supported by the Australian Research Council Centre of Excellence for All Sky Astrophysics in 3 Dimensions (ASTRO 3D), through project number CE170100013.
	
	This project has received funding from the European Research Council (ERC) under the European Union's Horizon 2020 research and innovation programme (grant agreement no.\ 679627, project name FORNAX).
	
	JMvdH acknowledges support from the European Research Council under the European Union's Seventh Framework Programme (FP/2007--2013)/ERC Grant Agreement no.\ 291531 (HIStoryNU).
	
	This research has made use of the NASA/IPAC Extragalactic Database (NED), which is operated by the Jet Propulsion Laboratory, California Institute of Technology, under contract with the National Aeronautics and Space Administration.

	\section*{Data Availability}
	
	The source code of \sofia{}~2 is available from GitHub at \url{https://github.com/SoFiA-Admin/SoFiA-2/}. The source code of \sofia{}-X is available from GitHub at \url{https://github.com/AusSRC/SoFiAX/}. The WALLABY pre-pilot \hi{} data of the Eridanus cluster are available from the CSIRO ASKAP Science Data Archive (CASDA) at \url{http://hdl.handle.net/102.100.100/52548} (DOI: \href{https://doi.org/10.25919/0yc5-f769}{10.25919/0yc5-f769}). The WALLABY pilot \hi{} data of the Hydra cluster are available from CASDA at \url{http://hdl.handle.net/102.100.100/319460} (DOI: \href{https://doi.org/10.25919/0v41-1055}{10.25919/0v41-1055}).

	
	\bibliographystyle{mnras}

\begin{thebibliography}{99}
		\bibitem[\protect\citeauthoryear{Bonnarel et al.}{2011}]{Bonnarel2011}
			Allen R.~J., Ekers R.~D., Terlouw J.~P., Vogelaar M.~G.~R., 2011, Astrophysics Source Code Library, record ascl:1109.018
		\bibitem[\protect\citeauthoryear{Amdahl}{1967}]{Amdahl1967}
			 Amdahl G.~M., 1967, Validity of the single-processor approach to achieve large scale computing capabilities. AFIPS Joint Spring Conference Proceedings 30 (Atlantic City, NJ, April 18--20), AFIPS Press, Reston VA, pp~483--485
		\bibitem[\protect\citeauthoryear{Blue Bird et al.}{2020}]{BlueBird2020}
			Blue Bird J., Davis J., Luber N., van Gorkom J.~H., Wilcots E., et al., 2020, MNRAS, 492, 153
		\bibitem[\protect\citeauthoryear{Allen et al.}{2011}]{Allen2011}
			Bonnarel F., Fernique P., Bienaym\'{e} O., Egret D., Genova F., Louys M., Ochsenbein F., Wenger M., Bartlett J.~G., 2000, \aaps, 143, 33
		\bibitem[\protect\citeauthoryear{Calabretta}{2011}]{Calabretta2011}
			Calabretta M.~R., 2011, Astrophysics Source Code Library, record ascl:1108.003
		\bibitem[\protect\citeauthoryear{Camilo}{2018}]{Camilo2018}
			Camilo F., 2018, Nature Astronomy, 2, 594
		\bibitem[\protect\citeauthoryear{Dagum \& Menon}{1998}]{Dagum1998}
			Dagum L., Menon R., 1998, Computational Science \& Engineering, 1, 46
		\bibitem[\protect\citeauthoryear{de Blok et al.}{2018}]{deBlok2018}
			de Blok W.~J.~G., Walter F., Ferguson A.~M.~N., Bernard E.~J., van der Hulst J.~M., et al., 2018, ApJ, 865, 26
		\bibitem[\protect\citeauthoryear{Dewdney et al.}{2009}]{Dewdney2009}
			Dewdney P.~E., Hall P.~J., Schilizzi R.~T., Lazio T.~J.~L.~W., 2009, IEEE Proc., 97, 1482
		\bibitem[\protect\citeauthoryear{Dickinson et al.}{2004}]{Dickinson2004}
			Dickinson M., Stern D., Giavalisco M., Ferguson H.~C., Tsvetanov Z., et al., 2004, ApJ, 600, L99
		\bibitem[\protect\citeauthoryear{Fl\"{o}er \& Winkel}{2012}]{Floeer2012}
			Fl\"{o}er L., Winkel B., 2012, PASA, 29, 244
		\bibitem[\protect\citeauthoryear{For et al.}{2019}]{For2019}
			For B.-Q., Staveley-Smith L., Westmeier T., Whiting M., Oh S.-H., et al., 2019, MNRAS, 489, 5723
		\bibitem[\protect\citeauthoryear{Guzman et al.}{2019}]{Guzman2019}
			Guzman J., Whiting M., Voronkov M., Mitchell D., Ord S., et al., 2019, Astrophysics Source Code Library, record ascl:1912.003
		\bibitem[\protect\citeauthoryear{Ho}{2007}]{Ho2007}
			Ho L.~C., 2007, ApJ, 669, 821
		\bibitem[\protect\citeauthoryear{Hotan et al.}{2021}]{Hotan2021}
			Hotan A.~W., Bunton J.~D., Chippendale A.~P., Whiting M., Tuthill J., et al., 2021, PASA, in press
		\bibitem[\protect\citeauthoryear{Jonas}{2009}]{Jonas2009}
			Jonas J.~L., 2009, IEEE Proc., 97, 1522
		\bibitem[\protect\citeauthoryear{Jurek}{2012}]{Jurek2012}
			Jurek R.~J., 2012, PASA, 29, 251
		\bibitem[\protect\citeauthoryear{Koribalski et al.}{2020}]{Koribalski2020}
			Koribalski B.~S., Staveley-Smith L., Westmeier T., Serra P., Spekkens K., et al., 2020, Ap\&SS, 365, 118
		\bibitem[\protect\citeauthoryear{Kova\v{c}, Oosterloo \& van der Hulst}{2009}]{Kovac2009}
			Kova\v{c} K., Oosterloo T.~A., van der Hulst J.~M., 2009, MNRAS, 400, 743
		\bibitem[\protect\citeauthoryear{Pence et al.}{2010}]{Pence2010}
			Pence W.~D., Chiappetti L., Page C.~G., Shaw R.~A., Stobie E., 2010, A\&A, 524, A42
		\bibitem[\protect\citeauthoryear{Popping et al.}{2012}]{Popping2012}
			Popping A., Jurek R.~J., Westmeier T., Serra P., Fl\"{o}er L., Meyer M., Koribalski B.~S., 2012, PASA, 29, 318
		\bibitem[\protect\citeauthoryear{Serra et al.}{2012a}]{Serra2012a}
			Serra P., Jurek R.~J., Fl\"{o}er L., 2012a, PASA, 29, 296
		\bibitem[\protect\citeauthoryear{Serra et al.}{2012b}]{Serra2012b}
			Serra P., Oosterloo T., Morganti R., Alatalo K., Blitz L., et al., 2012b, MNRAS, 422, 1835
		\bibitem[\protect\citeauthoryear{Serra et al.}{2015}]{Serra2015}
			Serra P., Westmeier T., Giese N., Jurek R.~J., Fl\"{o}er L., et al., 2015, MNRAS, 448, 1922
		\bibitem[\protect\citeauthoryear{Swaters et al.}{2002}]{Swaters2002}
			Swaters R.~A., van Albada T.~S., van der Hulst J.~M., Sancisi R., 2002, A\&A, 390, 829
		\bibitem[\protect\citeauthoryear{Taylor}{2005}]{Taylor2005}
			Taylor M.~B., 2005, in Shopbell P., Britton M., Ebert R., eds, ASP Conf.\ Ser., vol.~347, Astronomical Data Analysis Software and Systems XIV, p.~29
		\bibitem[\protect\citeauthoryear{Tully \& Fouqu\'{e}}{1985}]{Tully1985}
			Tully R.~B., Fouqu\'{e} P., 1985, ApJS, 58, 67
		\bibitem[\protect\citeauthoryear{Verheijen et al.}{2008}]{Verheijen2008}
			Verheijen M.~A.~W., Oosterloo T.~A., van Cappellen W.~A., Bakker L., Ivashina M.~V., van der Hulst J.~M., 2008, in Minchin R., Momjian E., eds, AIP Conf. Ser., vol.~1035, The Evolution of Galaxies Through the Neutral Hydrogen Window, p.~265
		\bibitem[\protect\citeauthoryear{Westmeier et al.}{2014}]{Westmeier2014}
			Westmeier T., Jurek R., Obreschkow D., Koribalski B.~S., Staveley-Smith L., 2014, MNRAS, 438, 1176
		\bibitem[\protect\citeauthoryear{Whiting}{2012}]{Whiting2012}
			Whiting M.~T., 2012, MNRAS, 421, 3242
		\bibitem[\protect\citeauthoryear{Whiting \& Humphreys}{2012}]{Whiting2012b}
			Whiting M.~T., Humphreys B., 2012, PASA, 29, 371
		\bibitem[\protect\citeauthoryear{Yan \& Windhorst}{2004}]{Yan2004}
			Yan H., Windhorst R.~A., 2004, ApJ, 612, L93
		\bibitem[\protect\citeauthoryear{Yoo, Jette \& Grondona}{2003}]{Yoo2003}
			Yoo A.~B., Jette M.~A., Grondona M., 2003, in Feitelson D., Rudolph L., Schwiegelshohn U., eds, Lecture Notes in Computer Science, vol.~2862, Job Scheduling Strategies for Parallel Processing, JSSPP 2003, Springer, Berlin, Heidelberg, p.~44
	\end{thebibliography}

	\appendix
	
	\section{Signal-to-noise ratio}
	\label{sect_snr}
	
	For the analysis presented in this work we have made use of $\mathrm{SNR}_{99}$ to characterise the signal-to-noise ratio of mock galaxies. For this purpose we first convolve the source model by the actual telescope beam used in the simulation. We then derive $\mathrm{SNR}_{99}$ by summing the flux densities in the individual pixels making up the source in the order of decreasing flux density up to the point where 99\% of the total flux has been accumulated. $\mathrm{SNR}_{99}$ is then obtained by dividing that summed flux density, $S_{99}$, by the associated statistical uncertainty as defined in Eq.~\ref{eqn_flux_unc}, assuming the actual noise level per pixel, $\sigma_{\rm rms}$, from the final mock data cube. Combining Eq.~\ref{eqn_flux} and~\ref{eqn_flux_unc} thus yields
	\begin{equation}
		\mathrm{SNR}_{99} = \frac{S_{99}}{\sqrt{N_{99} \, \Omega} \, \sigma_{\rm rms}} 
	\end{equation}
	where $\Omega$ is the beam solid angle in units of pixels and $N_{99}$ is the actual number of pixels that had to be summed to obtain $S_{99}$.
	
\begin{figure}
		\includegraphics[width=\linewidth]{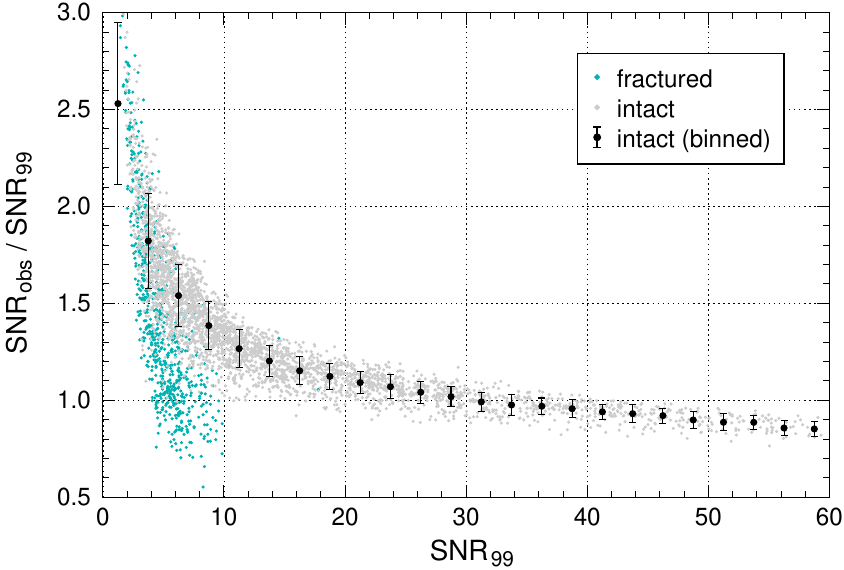}
		\caption{Ratio of observed signal-to-noise ratio and $\mathrm{SNR}_{99}$ as a function of $\mathrm{SNR}_{99}$. The meaning of the different colours and markers is the same as in Fig.~\ref{fig_param}.}
		\label{fig_snr99}
	\end{figure}

	An important question that arises from the use of $\mathrm{SNR}_{99}$ is how it relates to the observed signal-to-noise ratio, $\mathrm{SNR}_{\rm obs}$, of a source, i.e.\ the signal-to-noise ratio that arises from summing over the actual source mask obtained with \sofia{}~2. The ratio of $\mathrm{SNR}_{\rm obs} / \mathrm{SNR}_{99}$ of the sources detected in the mock data experiment in Section~\ref{sect_performance} is plotted in Fig.~\ref{fig_snr99} as a function of $\mathrm{SNR}_{99}$. As can be seen, the two are roughly equal at a signal-to-noise ratio of about~30. Towards the upper end of the SNR range, $\mathrm{SNR}_{\rm obs}$ becomes smaller than $\mathrm{SNR}_{99}$, suggesting that \sofia{}~2 is summing over more pixels than necessary to obtain 99\% of the flux of a source. The sharp rise in $\mathrm{SNR}_{\rm obs} / \mathrm{SNR}_{99}$ towards the lower end of the SNR range is largely the result of noise bias and likely related to the the peak flux density bias shown in the upper-left panels of Fig.~\ref{fig_param} and~\ref{fig_param2}.
	
	The main conclusion from our analysis is that for faint sources the observed integrated signal-to-noise ratio is somewhat larger than $\mathrm{SNR}_{99}$, presumably due to a combination of the aforementioned noise bias and the fact that the source mask might only enclose the brighter parts of the source, thus elevating its signal-no-noise ratio. Hence, any specific completeness limit in $\mathrm{SNR}_{99}$ space is likely to correspond to a somewhat higher limit in $\mathrm{SNR}_{\rm obs}$ space.
	
	In principle it should be possible to derive a quantitative relationship between $\mathrm{SNR}_{99}$ and $\mathrm{SNR}_{\rm obs}$ from the analysis presented here, but in practice such a relation would be of little use in our experience, as the actual $\mathrm{SNR}_{\rm obs}$ measurement will strongly depend on both the settings used in \sofia{}~2 and the characteristics of the data set to be searched, and for those reasons cannot be generalised. As a basic example, choosing a higher detection threshold in the S+C finder would result in a more compact source mask and thus a larger value of $\mathrm{SNR}_{\rm obs}$, whereas enabling the mask dilation algorithm would grow the size of the mask and thus result in a smaller value of $\mathrm{SNR}_{\rm obs}$ for the same source. This is because the noise scales with the square root of the number of pixels in the mask, while the signal can to first order be treated as constant.

	\bsp
	\label{lastpage}
\end{document}